\documentclass[twocolumn]{aa}  

\usepackage{natbib,twoopt}
\usepackage[breaklinks=true]{hyperref} 
\bibpunct{(}{)}{;}{a}{}{,}             
\makeatletter
  \newcommandtwoopt{\citeads}[3][][]{\href{http://adsabs.harvard.edu/abs/#3}%
    {\def\hyper@linkstart##1##2{}%
     \let\hyper@linkend\@empty\citealp[#1][#2]{#3}}}
  \newcommandtwoopt{\citepads}[3][][]{\href{http://adsabs.harvard.edu/abs/#3}%
    {\def\hyper@linkstart##1##2{}%
     \let\hyper@linkend\@empty\citep[#1][#2]{#3}}}
  \newcommandtwoopt{\citetads}[3][][]{\href{http://adsabs.harvard.edu/abs/#3}%
    {\def\hyper@linkstart##1##2{}%
     \let\hyper@linkend\@empty\citet[#1][#2]{#3}}}
  \newcommandtwoopt{\citeyearads}[3][][]%
    {\href{http://adsabs.harvard.edu/abs/#3}
    {\def\hyper@linkstart##1##2{}%
     \let\hyper@linkend\@empty\citeyear[#1][#2]{#3}}}
\makeatother

\usepackage{color}
\usepackage{xcolor}
\usepackage{txfonts}
\usepackage[version=4]{mhchem}
\usepackage{placeins}
\usepackage{multirow}
\usepackage{verbatim}

\begin{document}
\title{Probing planet formation and disk substructures in the inner disk of Herbig Ae stars with CO rovibrational emission}
\author{Arthur D. Bosman \inst{1} \and Andrea Banzatti \inst{2,3} \and Simon Bruderer \inst{4} \and Alexander G. G. M. Tielens \inst{1} \and Geoffrey A. Blake \inst{5} \and Ewine F. van Dishoeck \inst{1,4} }
\institute{Leiden Observatory, Leiden University, PO Box 9513, 2300 RA Leiden, The Netherlands\\ \email{bosman@strw.leidenuniv.nl},
\and
Department of Physics, Texas State University, 749 N Comanche Street, San Marcos, TX 78666, USA
\and
Department of Planetary Sciences, University of Arizona, 1629 East University Boulevard, Tucson, AZ 85721, USA 
\and
Max-Planck-Institut f\"{u}r Extraterrestrische Physik, Gie{\ss}enbachstrasse 1, 85748 Garching, Germany
\and Division of Geological \& Planetary Sciences, California Institute of Technology, 1200 E California Blvd, Pasadena, CA 91125}
\abstract 
{CO rovibrational lines are efficient probes of warm molecular gas and can give unique insights into the inner 10 AU of proto-planetary disks, effectively complementing ALMA observations. Recent studies have found a relation between the ratio of lines originating from the second and first vibrationally excited state, denoted as $v2/v1$, and the Keplerian velocity or emitting radius of CO. 
Counterintuitively, in disks around Herbig Ae stars the vibrational excitation is low when CO lines come from close to the star, and high when lines only probe gas at large radii (more than 5 AU). The $v2/v1$ ratio is also counterintuitively anti-correlated with the near-IR (NIR) excess, which probes hot/warm dust in the inner disk.}
{We aim to find explanations for the observed trends between CO vibrational ratio, emitting radii, and NIR excess, and identify their implications in terms of the physical and chemical structure of inner disks around Herbig stars. }
{First, slab model explorations in LTE and non-LTE are used to identify the essential parameter space regions that can produce the observed CO emission. Second, we explore a grid of thermo-chemical models using the DALI code, varying gas-to-dust ratio and inner disk radius. Line flux, line ratios and emitting radii are extracted from the simulated lines in the same way as the observations and directly compared to the data. }
{Broad CO lines with low vibrational ratios are best explained by a warm (400--1300 K) inner disk surface with gas-to-dust ratios below $1000$ ($N_\mathrm{CO} < 10^{18}$ cm$^{-2}$); no CO is detected within/at the inner dust rim, due to dissociation at high temperatures. In contrast, explaining the narrow lines with high vibrational ratios requires an inner cavity of a least 5 AU in both dust and gas, followed by a cool (100--300 K) molecular gas reservoir with gas-to-dust ratios greater than 10000 ($N_\mathrm{CO} > 10^{18}$ cm$^{-2}$) at the cavity wall. In all cases the CO gas must be close to thermalization with the dust ($T_\mathrm{gas} \sim T_\mathrm{dust}$).}
{The high gas-to-dust ratios needed to explain high $v2/v1$ in narrow CO lines for a subset of group I disks can naturally be interpreted as due to the dust traps that have been proposed to explain millimeter dust cavities. The dust trap and the low gas surface density inside the cavity are consistent with the presence of one or more massive planets. The difference between group I disks with low and high NIR excess can be explained by gap opening mechanisms that do or do not create an efficient dust trap, respectively. The broad lines seen in most group II objects indicate a very flat disk in addition to inner disk substructures within 10 AU that can be related to the substructures recently observed with ALMA. We provide simulated ELT-METIS images to directly test these scenarios in the future. 
}  
\titlerunning{Probing the inner disk of Herbig Ae sources with CO rovibrational emission}
\keywords{protoplanetary disks -- molecular processes -- astrochemistry -- radiative transfer -- line: formation}
\maketitle

\section{Introduction}

\begin{figure*}
\centering
\begin{minipage}{0.5\hsize}
\centering
\includegraphics[width = 0.65\hsize]{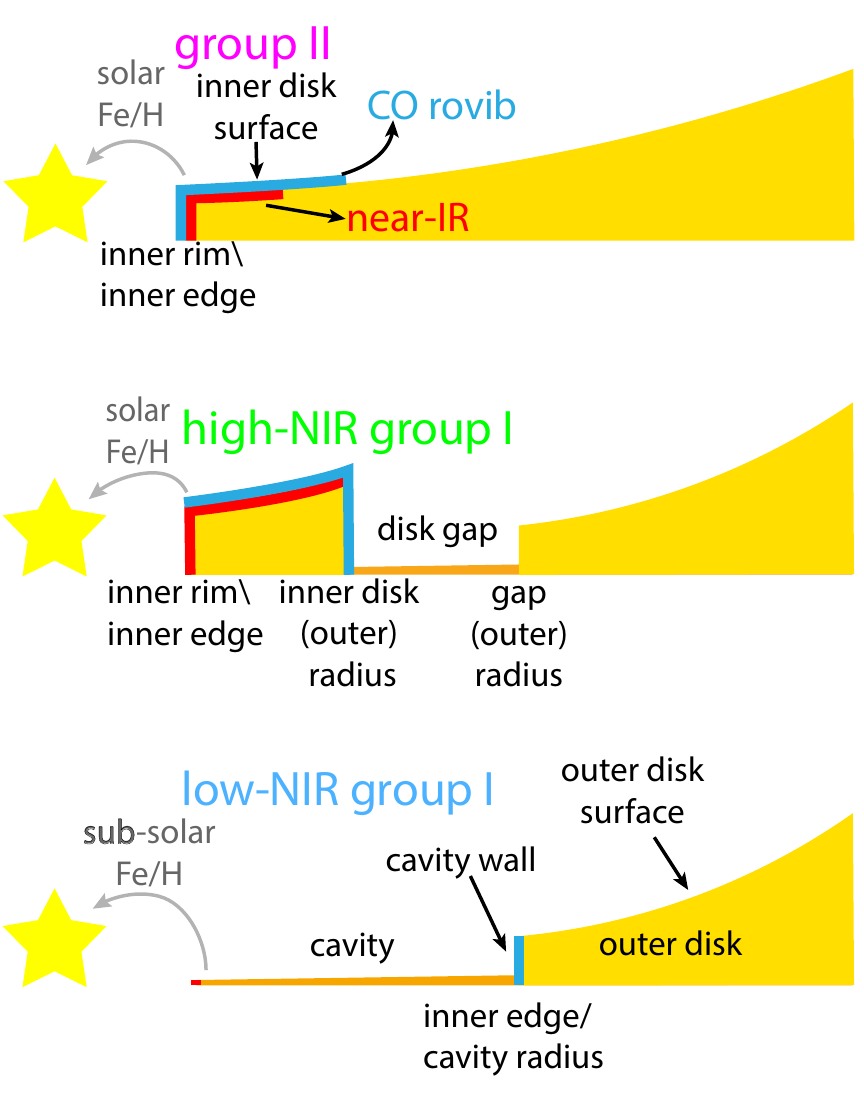}
\end{minipage}%
\begin{minipage}{0.5\hsize}
\includegraphics[width = \hsize]{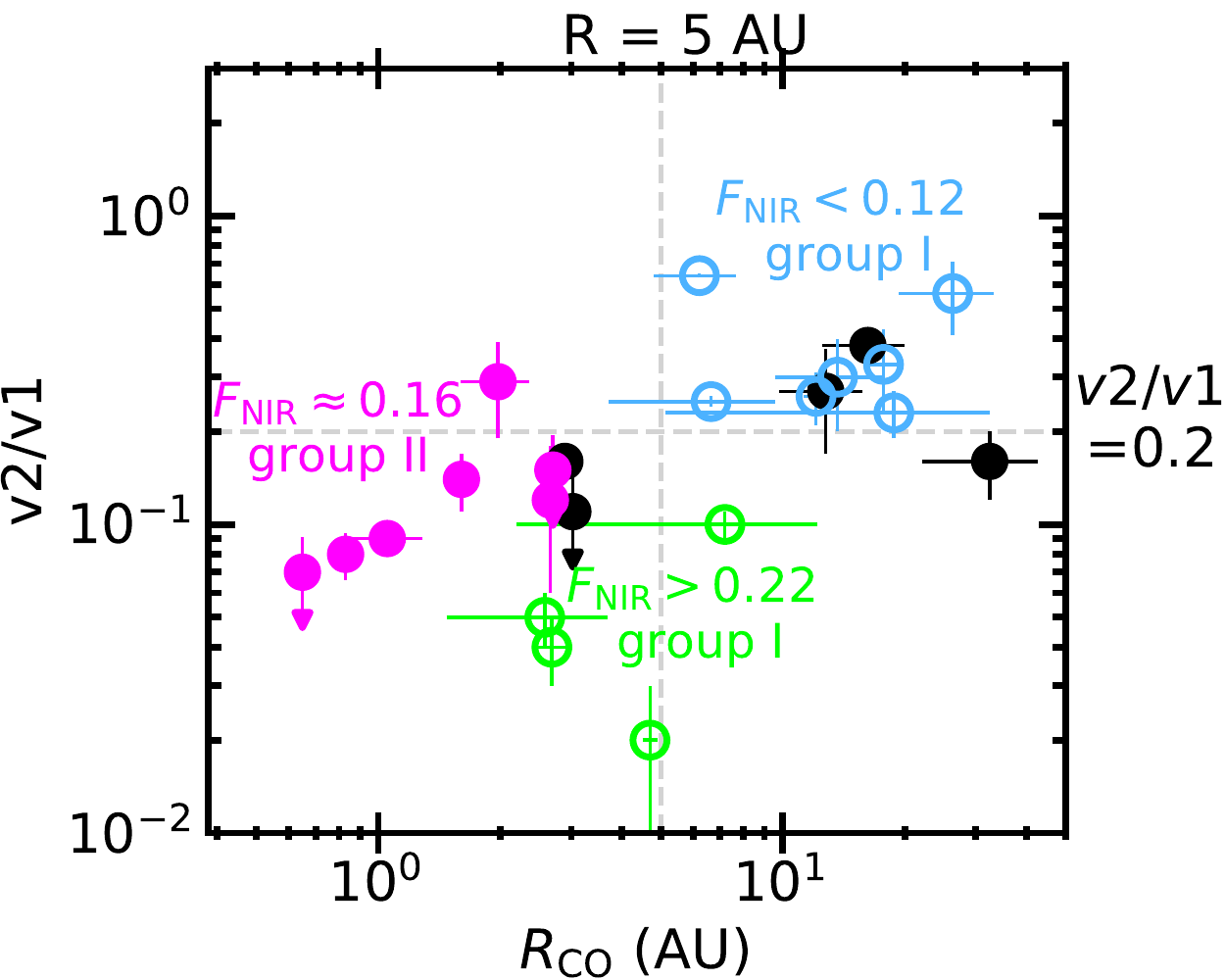}
\end{minipage}
\caption{\label{fig:data}
Disk structures as proposed by \citet{Banzatti2018} (left) and 
CO vibrational ratio $v2/v1$ and emitting radius from near-infrared CO spectra of Herbig stars used for comparison to the models in this work (right), see details in Section \ref{sec:data}). The three groups from \citet{Banzatti2018}, are shown in different colors: group II in magenta, high-NIR group I in green, low-NIR group I in blue. Disks where $F_\mathrm{NIR}$ is not available are marked in black. We mark two regions that will be used for comparison with models: disks that have CO inside 5 AU, which show a low vibrational ratio (group II and high-NIR group I, \textit{bottom left corner}), and disks that have CO only outside of 5 AU, which show high vibrational ratios (low-NIR group I, \textit{top right corner}) .
}

\end{figure*}

\begin{table*}[]
\centering
    \caption{Median values for Herbig groups from \citet{Banzatti2018}, plus notes from imaging studies}
    \centering
    
    \begin{tabular}{l c c c c l l}
    \hline
    \hline
    Herbig group & $F_\mathrm{NIR}$ & $v2/v1$ & $R_\mathrm{CO}$ (AU) & log(Fe/H) & Dust structures from imaging & Refs \\
    \hline
   Group II& 0.16 & 0.12 & 3 & -4.4 & no large inner cavities, some substructures $< 5$~AU & (1) \\
   Group I (high-NIR) & 0.27 & 0.05 & 5 & -4.6 & 40--100~AU cavities, (misaligned) inner disks $< 10$~AU & (2) \\
   Group I (low-NIR) & 0.08 & 0.27 & 18 & -5.2 & 15--50~AU cavities, no significant inner disks & (3) \\
    \hline
    \end{tabular}
    \tablefoot{References: (1) e.g. \citet{Menu2015, Huang2018, Isella2018} ; (2) e.g. \citet{Pinilla2018, Stolker2016, Avenhaus2017, Tang2017, diFolco2009, Boehler2018, Benisty2017}; (3) \citet{vanderPlas2017,Fedele2017, White2018, Pinilla2018} }
    \label{tab:Av_vals}
\end{table*}

Planetary systems are thought to be built within proto-planetary disks of gas and dust around young stars. How these disks transition from the initial gas-rich remnants of star formation to the solid-body dominated debris disks and planetary systems is still an open question. While for most disks it seems that they go through a quick dispersal process \citep{Cieza2007, Currie2009}, there is a subset of disks that goes through a prolonged period where dust and gas are (partially) depleted in the inner disk, but where there is a large reservoir of mass at larger radii, the so called transition disks \citep{Maaskant2013, Garufi2017, vanderMarel2018}. It is thought that in these disks the cavity is formed either through giant planet formation or X-ray/UV photoevaporation \citep[for reviews, see][]{Owen2016, Ercolano2017}. These processes cause different distributions of gas and dust in the inner disk.

To distinguish these scenarios, we have to study the inner disk. CO rovibrational lines are good tracers of the inner disk, because they are strong lines that originate only from warm, dense gas. The upper level energies for the first vibrationally excited level are around 3000 K so they will only be excited in environments with high temperatures ($\gtrsim$ 300 K). Furthermore CO is a very stable molecule and is thus expected to survive even in regions where there is little dust to shield the gas from UV photons \citep[e.g.][]{Bruderer2013}. Finally the transitions are strong so small columns of excited CO are needed to produce bright lines making CO columns as low as $10^{16}$ cm$^{-2}$ easily detectable if the excitation conditions are right. 

Observing the fundamental CO lines around 4.7 $\mu$m allows for the simultaneous measurement of rovibrational line fluxes from the first ($v1$) and second ($v2$) excited states. The $v2/v1$ line flux ratio carries information on the excitation conditions of the gas. The high resolving power on spectrographs such as Keck-NIRSPEC \citep{McLean1998}, VLT-CRIRES \citep{Kaufl2004}, IRTF-CSHELL and now iSHELL \citep{Rayner2016} make velocity-resolved observations of CO line profiles possible \citep[e.g.][]{Najita2003, Blake2004, Thi2005, Brittain2007, Pontoppidan2008, SalykCO2011, Brown2013, Banzatti2015, Brittain2018}. As the emission is expected to come from a Keplerian disk, the width of the line, once coupled with disk inclination and stellar mass, can be used to estimate the CO emitting radii for different gas velocities, thereby obtaining information on the spatial distribution of CO gas in inner disks. 

For disks around T-Tauri stars, CO rovibrational lines have been used to probe molecular gas within inner disk dust cavities \citep{Pontoppidan2008, Pontoppidan2011}, and to propose an inside-out clearing scenario for gas and dust \citep{Banzatti2015}. In their sample of T-Tauri disks, CO shows a lower vibrational temperature with decreasing linewidth (and hence increasing emitting radius). This fits well with the expectation that the gas temperature decreases as a function of distance from the star. Furthermore the variation in CO emitting radii can be explained by varying inner molecular gas disk radii, indicating that the inner edge of the molecular disk is not set by the sublimation of dust but carved by another process such as planet formation or photoevaporation.

\begin{figure*}
    \centering
    \includegraphics[width = \hsize]{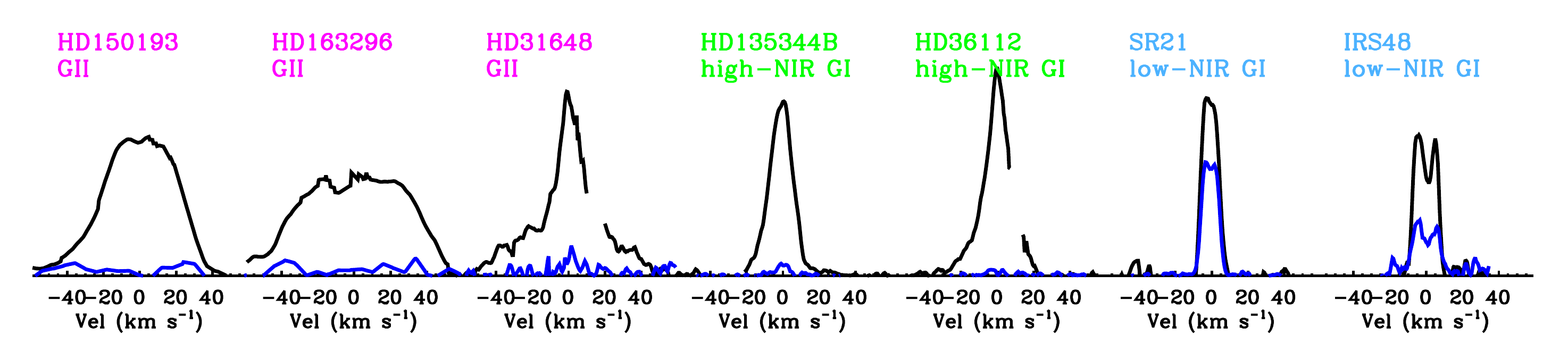}
    \caption{Selection of stacked CO line profiles from observed spectra (Section \ref{sec:data}). The $v1 $ lines are shown in black, $v2 $ lines in blue. Gaps visible in some line profiles are due to telluric absorption. Disk inclinations are between 20 and 50 deg for all these objects. In HD 31648, the $R_\mathrm{CO}$ is taken for the broad component defined by the line wings.
    }
    \label{fig:obslineprofiles}
\end{figure*}

Herbig disks, instead, behave very differently and show an inverse relation between linewidth and vibrational excitation \citep{Banzatti2015}. This was attributed to UV-fluorescence becoming more important for stars with higher continuum UV fluxes when the thermal excitation becomes less efficient at larger disk radii \citep{Brittain2007, Brown2013, Banzatti2015}. However, full thermo-chemical models suggest that UV fluorescence is not the dominant excitation mechanism for the $v=1$ and $v=2$ levels of CO \citep{Thi2013, HeinBertelsen2014}. This is supported by the observed rovibrational excitation diagrams that only show a strong difference between rotational and vibrational temperatures for levels $v =3$ and higher, indicating that only for the higher vibrational levels, UV pumping is important \citep{vanderPlas2015}. 

Furthermore, based on their SEDs, Herbigs are divided into two groups. Disks with strong far-infrared emission relative to their mid-infrared emission are classified as group I or "flared" disks. Disks with strong mid-infrared emission in comparison to their far infrared emission are classified as group II or "settled" disks \citep{Meeus2001}. 
An evolutionary sequence between these groups was inferred with group I disk being the precursors of group II disks \citep{Dullemond2004}. However it is the group I disk that are often observed to have a large ($> 10$ AU) cavity in either scattered light or sub-mm imaging, implying that they are unlikely the precursors for the group II disks that are generally less massive and do not show any cavity \citep{Maaskant2013, Garufi2017}.

Inner disk ($\lesssim 5$ AU) tracers of both gas and dust add interesting pieces to this puzzle. Table~\ref{tab:Av_vals} reports median values for three inner disk tracers (near-infrared excess, $F_\mathrm{NIR}$; CO vibrational ratio, $v2/v1$; CO emitting radius, $R_{\ce{CO}}$) as well as the median stellar surface Fe abundance \citep{Kama2015} used to identify three groups of Herbig disks in \cite{Banzatti2018}; here we also add notes on the presence of disk structures from imaging studies. 
Group II disks exhibit a narrow range of intermediate values for the near-infrared excess, $F_\mathrm{NIR} = L_{1.2-4.5 \mu\mathrm{m}}/L_\star$ \citep{Garufi2017}. All of the group II disks show broad CO 4.7 $\mu$m rovibrational lines indicating that CO is emitting from small radii. These tracers together indicate that both molecular gas and dust are present and abundant at small distances from the star \citep[$\lesssim 5$ AU;][]{vanderPlas2015, Banzatti2018}.

The group I disks, instead, remarkably split into two very distinct groups \citep{Banzatti2018}. Some of them have very high near-infrared excesses (high-NIR), higher than the group II sources, and only moderately broad CO rovibrational lines. The rest of the group I disks have low near-infrared excesses (low-NIR) and the narrowest CO rovibrational lines. Group I disks thus, while all having dust cavities imaged at larger radii, seem to show a marked dichotomy in their inner disks between those that have abundant gas and dust in the inner few AU, and those that do not, without a gradient of situations in between. While these groups do not show any segregation in terms of mass accretion rates \citep{Banzatti2018}, stellar elemental abundances show that the low-NIR group I disks are depleted in Fe compared to all of the other sources (Table~\ref{tab:Av_vals}), suggesting that the stars in low-NIR group I disks accrete gas that is depleted in dust compared to the 100:1 ISM dust ratio, suggesting that dust is trapped at larger radii in the disk \citep{Kama2015}. 

In this work we focus on CO rovibrational emission, and in particular the observed trends between the radius and excitation of CO emission and the NIR excess (Fig.~\ref{fig:data}), to expand our growing understanding of inner disk structure and evolution in Herbigs. 
Specifically, we aim to explain the dichotomy between low vibrational ratios coming from gas within $< 5$ AU and the high vibrational ratios coming from larger radii. The observational dataset from \cite{Banzatti2017,Banzatti2018}, briefly presented in Sec.~\ref{sec:data}, is used for comparison and validation of the models. In Sec.~\ref{sec:Slabmodelling} the vibrational excitation of CO will be studied through simple slab models. Full thermo-chemical models using Dust And LInes \citep[DALI,][]{Bruderer2012,Bruderer2013} for different physical structures will be presented and analysed in Sec.~\ref{sec:DALI}. The implications will be discussed in Sec.~\ref{sec:discussion} and our conclusion will be summarized in Sec.~\ref{sec:conclusion}. 

\section{Data overview}
\label{sec:data}

The CO emission lines adopted in this work for comparison to the models are taken from the compilation included in \citet{Banzatti2017,Banzatti2018}, based on spectra originally presented in \citet{Pontoppidan2011b,Brown2012,Banzatti2015a, vanderPlas2015,Banzatti2018}. The data consist of high resolution ($R \sim $75,000--100,000) spectra of CO rovibrational emission around 4.7 $\mu$m for 20 Herbig Ae stars and 3 F stars, taken with the CRIRES instrument on the Very Large Telescope (VLT) of the European Southern Observatory \citep[ESO; ][]{Kaufl2004} and iSHELL on the NASA Infrared Telescope Facility  \citep[IRTF; ][]{Rayner2016}. The spectrum of HD 142666 is taken from a previous survey \citep{Blake2004,SalykCO2011} done with Keck-NIRSPEC \citep[R $\sim$ 25,000;][]{McLean1998}. The two parameters we focus on in this work, the CO vibrational ratio, $v2/v1$, and a characteristic emitting radius, are measured from stacked line profiles as explained in \citet{Banzatti2015}.

In brief, the vibrational ratio $v2/v1$ is measured from the line flux ratio between lines around the $v2\,P(4)$ line ($v' = 2, J' = 3 \rightarrow v'' = 1, J'' = 4$) and around the $v1\,P(10)$ line ($v' = 1, J' = 9 \rightarrow v'' = 0, J'' = 10$). The choice of these specific lines is driven by the spectral coverage of the observations, and by the need to use unblended lines \citep[see details in][]{Banzatti2015}. The vibrational flux ratio between the $v2\,P(4)$ and $v1\,P(10)$ line is used as a proxy for the vibrational ratio between the $v2$ and $v1$ levels. 
The vibrational ratio depends on the lines that are used in the comparison, even lines of matching $J$ level show a variation of up to 50\% in the vibrational ratio. The $v2\,P(4)$ and $v1\,P(10)$ line ratio lies within the range of values obtained by using matching $J$ levels and is thus a good proxy for the vibrational ratio \citep[see Appendix A in][]{Banzatti2015}. 
A characteristic emitting radius is estimated from the half width at half maximum (HWHM) of the line profile, assuming Keplerian rotation and using literature values for the disk inclination and the stellar mass. As better measurements of disk inclinations have become available over time for some disks, estimates of CO radii have changed accordingly; the error-bars in Fig.~\ref{fig:data} reflect the uncertainties in the disk inclinations.
Figure~\ref{fig:data} shows these parameters and their trend as discussed above, namely that the vibrational ratio is larger when CO emission comes from larger disk radii.

Figure~\ref{fig:obslineprofiles} shows a selection of CO line profiles, chosen to span the full range of CO emitting radii and vibrational ratios for the three groups of disks in Fig.~\ref{fig:data}. The broader lines (i.e. smaller CO emitting radii) have low $v2/v1$, and can have flat or double-peaked line profiles. HD 31648 is the only exception that clearly shows two velocity components, as commonly found for T-Tauri stars \citep{Bast2011,Banzatti2015}. 
This combination of broad wings and strong peak indicates that the emitting area of the CO rovibrational lines spans a large range of radii (see more in Section \ref{sec:DALI}). In this analysis, for HD 31648 we take the CO radius as indicated by the broad component, defined by the broad line wings. The narrower lines (i.e. larger CO emitting radii) often show a single peak profile indicative of a more extended emitting area, but in some cases they clearly show a double peak profile, indicative of an emitting region that is confined to a narrower ring. 

In addition, we use CO line fluxes as measured in \citet{Banzatti2017}, which we scale to a common distance of 150 pc for comparison with the model. The near-infrared excess is measured between 1.2 and 4.5 $\mu$m \citep{Garufi2017,Banzatti2018}. Table \ref{tab:Av_vals} shows the median values for these parameters for the three groups of Herbig disks, as reported in \citet{Banzatti2018}. Spectra and individual measurements can be found in the original references reported in this section.

\section{Slab modelling of the vibrational ratio}
\label{sec:Slabmodelling}
To be able to infer the physical conditions of the CO emitting regions we have to look at the CO line formation process. To compute the strength of a line one needs to know both the chemical and physical state of the gas. Physics and chemistry are strongly intertwined with the temperature, density and radiation field influencing the chemistry and the chemical abundances influencing the heating and cooling of the gas, changing the temperature. Various thermo-chemical models have been developed it solve this coupled problem \citep[e.g.][]{Woitke2009, Bruderer2012, Bruderer2013, Du2014}, however, before we dive into the full problem, we will first study the line formation of CO in a more controlled setting. 

The line formation of CO will be studied using two different types of slab models. First the behaviour of a slab of CO with fixed excitation will be studied analytically; this will reveal the effects of the optical depth and excitation on the vibrational ratios. Afterwards RADEX models \citep{RADEX} will be used to study non-LTE effects. These RADEX models will be used to constrain the physical conditions of the CO rovibrational emitting regions. 

\subsection{Analytical line ratios}
\subsubsection{Methods}
\begin{figure}
\includegraphics[width=\hsize]{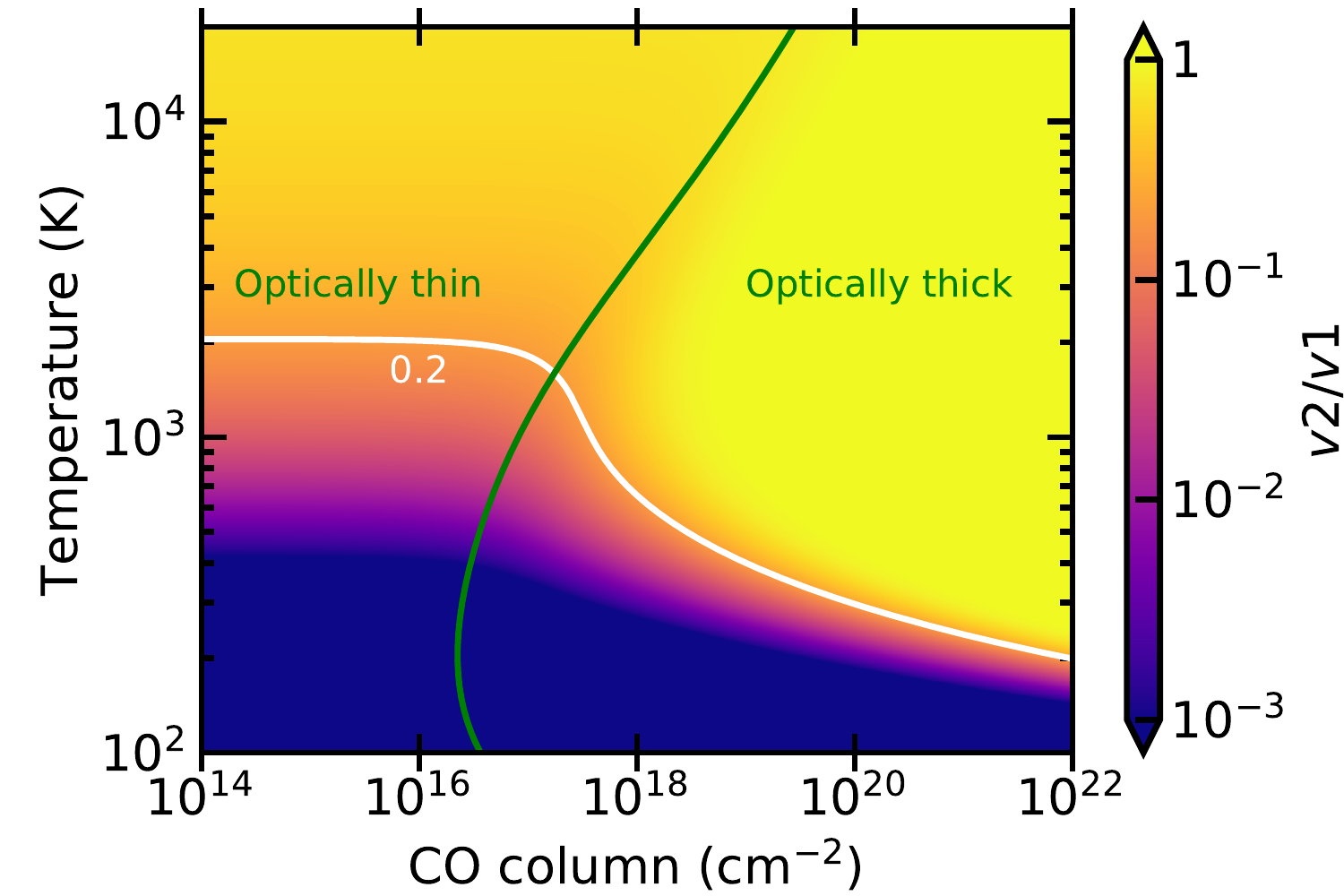}
\caption{\label{fig:Analytic_ratio} CO vibrational ratio, $v2/v1$, for different temperatures and columns from the analytic model. The green line shows the $\tau = 1$ conditions for the $v1$ line. The white line shows $v2/v1 = 0.2$, which is the value that differentiates low and high vibrational ratio sources.  }
\end{figure}

In the case of a mono-thermal slab of CO that is in LTE with an excitation temperature $T_{\mathrm{ex}}$, the continuum subtracted peak surface brightness can be computed by:
\begin{equation}
\label{eq:CO_I}
I(u, l) = \left(B\left((T_{\mathrm{ex}}, \nu_{(u, l)}\right) - B\left(T_\mathrm{back}, \nu_{(u, l)}\right)\right) \times \left( 1 - e^{-\tau(u, l)}\right),
\end{equation}
where
\begin{equation}
\label{eq:CO_tau}
\tau(u, l)  = \frac{g(u)}{g(l)} \frac{c^2A(u,l)}{8 \pi^2 \nu^2(u,l)}
 \frac{N_\mathrm{CO} \left(1 - \exp{-\left[\frac{h\nu(u,l)}{kT_{\mathrm{ex}}}\right]}\right)}{\sqrt{2\pi}\sigma_v Z(T_{\mathrm{ex}})} g(l)\exp{\left[-\frac{E(l)}{kT_{\mathrm{ex}}}\right]}.
\end{equation}
In Eq.~\ref{eq:CO_I} $I(u, l)$ is the continuum subtracted line peak intensity, $B(T, \nu)$ is the Planck function at temperature $T$ and frequency $\nu$, $T_\mathrm{back}$ is the radiation temperature of the background and $\tau(u, l)$ is the line peak opacity. In Eq.~\ref{eq:CO_tau} $g(n)$ is the degeneracy of rovibrational level $n$, $A(u,l)$ is the Einstein A coefficient of the transition between rovibrational levels $u$ and $l$, $N_\mathrm{CO}$ is the CO column, $\sigma_v$ is the thermal linewidth, $Z(T_\mathrm{ex})$ is the rovibrational partition function of CO, $E(n)$ is the energy above the rovibrational ground state of state $n$ and $c$, $h$ and $k$ are the speed of light, the Planck constant and the Boltzmann constant as usual. 

The $v1\,P(10)$ and the $v2\,P(4)$ lines are used as proxy for the stacked $v1$ and $v2$ line from the observations (Section \ref{sec:data}). Under the current assumptions the peak line intensity only depends on the excitation temperature, the total column and the background radiation temperature. This last parameter drops out when looking at line ratios (assuming that $T_\mathrm{back}$ does not vary significantly over the frequency range). 

\subsubsection{Results}
The peak line intensity ratio for the  $ v1 $ and $ v2$ lines are shown in  Fig.~\ref{fig:Analytic_ratio} for a range of temperatures and CO columns. At columns smaller than $10^{17}$ cm$^{-2}$ both lines are optically thin and as such there is no trend with column in the line ratio. At high column densities the line ratio converges to $g_2(u)A_2(u,l)/g_1(u)A_1(u,l)$ which is $\sim 1.01$ for the lines under consideration, for almost any temperature ($T >$ 200 K).

The green line shows where the $ v1 $ line becomes optically thick. An increase in CO column to the right of this line no longer elicits a linear response in the line flux. As the $ v2$ flux still increases linearly with the column, this increases the line ratio. If the column gets big enough the $ v2$ line also gets optically thick and the line ratio tends to unity. 
The speed at which this happens with increasing column strongly depends on the population of the upper level of the $ v2$ transition. At temperatures above, 2000 K the column at which the $ v1 $ becomes optically thick increases due to the lower fractional population in the lower rotational levels of the $ v1 $ line. 

Observed line ratios coming from within 5 AU are generally below 0.2, so from Fig.~\ref{fig:Analytic_ratio} the conditions to match these observations can be easily read off. If both lines are optically thin, an excitation temperature less than $\sim${}$2000$ K induces low line ratios. At columns above $10^{17}$ cm$^{-2}$ a line ratio of 0.2 requires lower temperature with increasing column, to values as low as $\sim${}$190$ K at a CO column of $10^{22}$ cm$^{-2}$. 

All disks with $R_\mathrm{CO} > 5$ AU have line ratios between 0.2 and 0.5. For these high line ratios Fig.~\ref{fig:Analytic_ratio} shows that, as expected a high line ratio can be due to high temperature, or large columns. For low columns, temperatures between $2000$ and $6000$ K are needed to produce the right line ratios. Above a column of $10^{17}$ cm$^{-2}$ progressively lower temperatures lead to the observed line flux ratios.

\begin{figure*}
\begin{minipage}{0.5\hsize}
\includegraphics[width=\hsize]{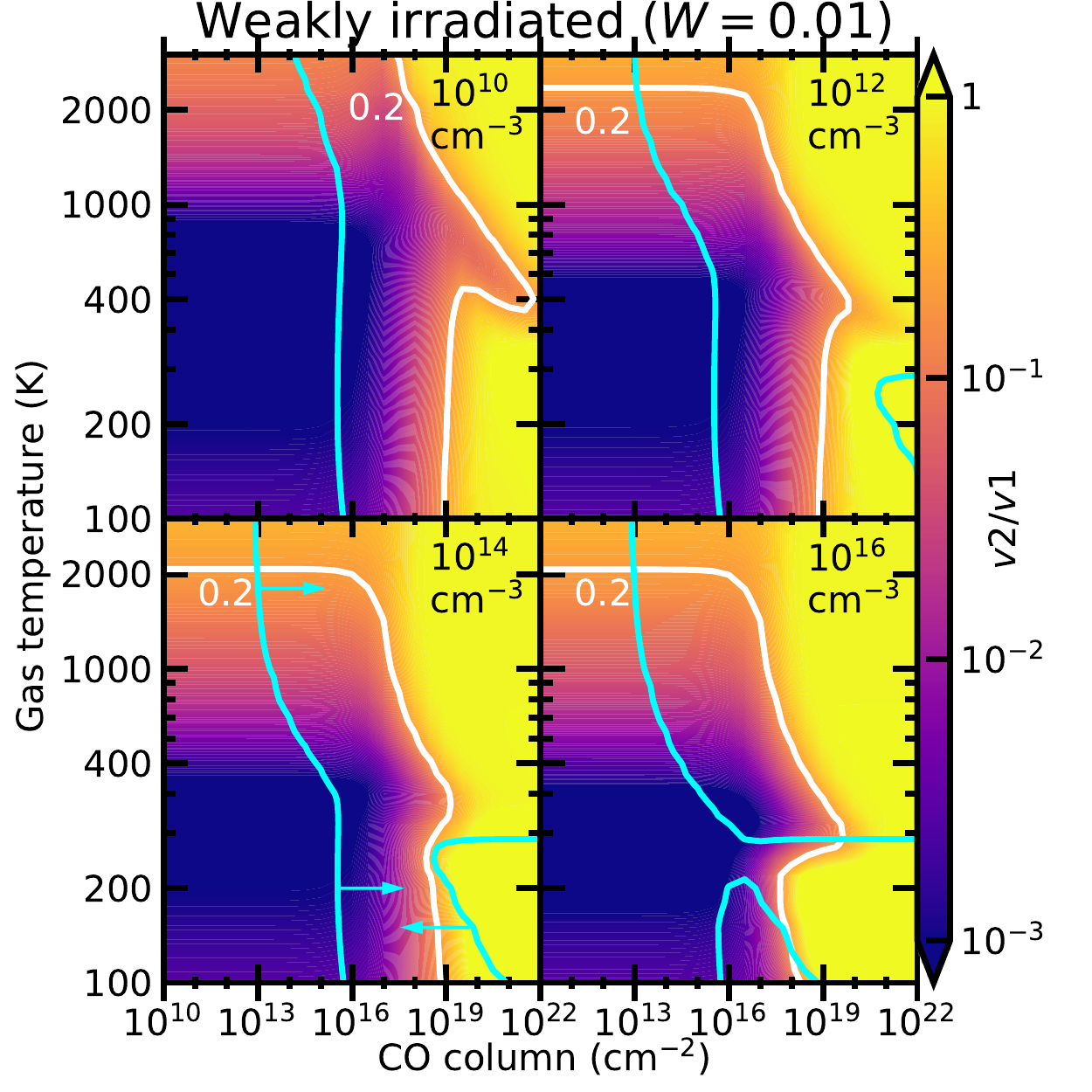}
\end{minipage}%
\begin{minipage}{0.5\hsize}
\includegraphics[width=\hsize]{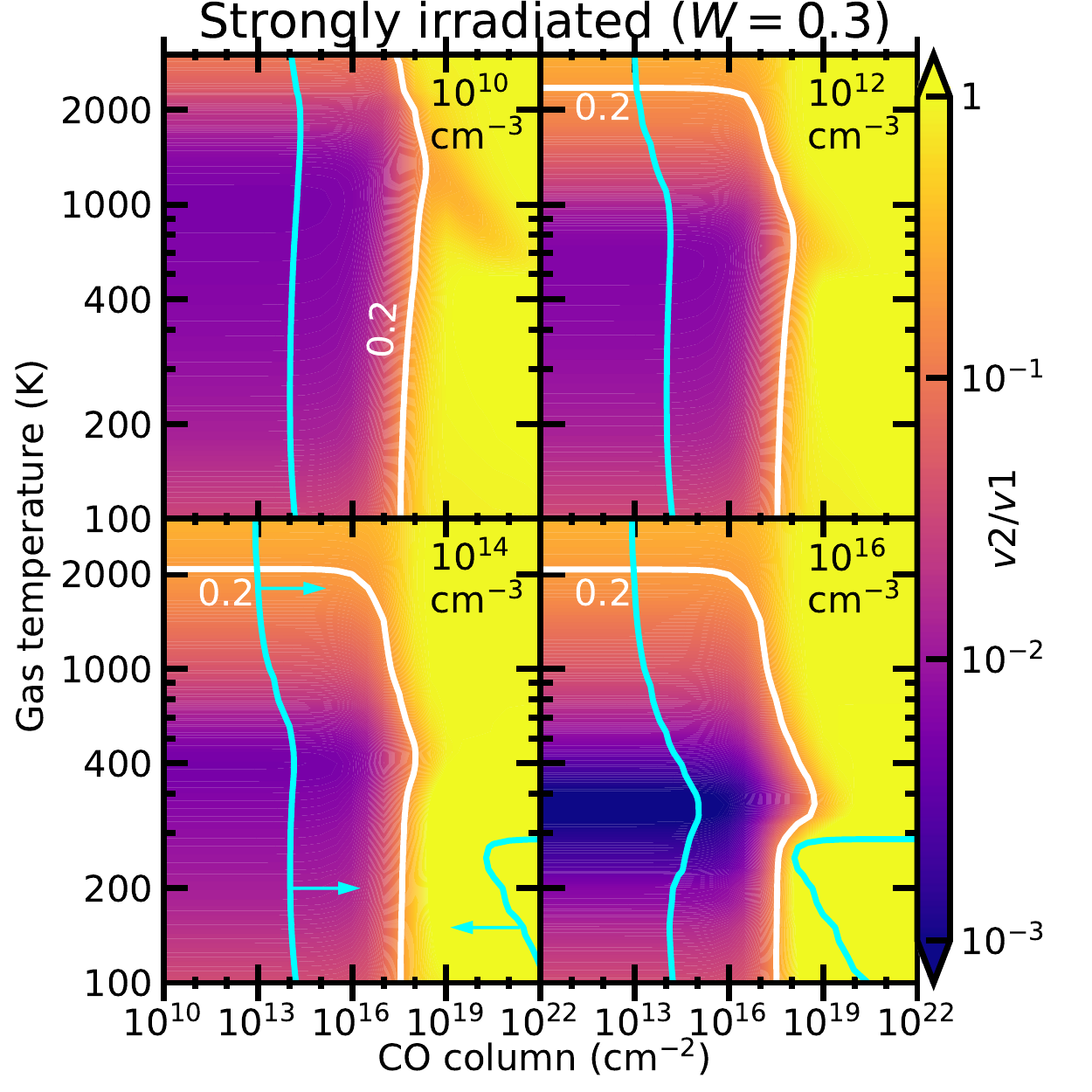}
\end{minipage}
\caption{\label{fig:RADEX_001_ratio}\label{fig:RADEX_03_ratio} CO vibrational ratio, $v2/v1$, for different temperatures and columns from the RADEX models using a 750 K radiation field with a dilution factor $W$ of 0.01 (\textit{left}) and 0.3 (\textit{right}). The area between the blue and white lines shows where both the vibrational ratio and the $v1$ flux of the low vibrational ratio sources are reproduced. For the $v1$ flux an emitting area with a radius of 5 AU is assumed. If a smaller emitting area is assumed the blue lines would shift in the direction of the blue arrows. High vibrational ratio sources can either be explained by gas with a high column ($N \gtrsim 10^{18}$ cm$^{-2}$) or a high temperature ($T > 2000$ K).}  
\end{figure*} 
\begin{figure}
\includegraphics[width = \hsize]{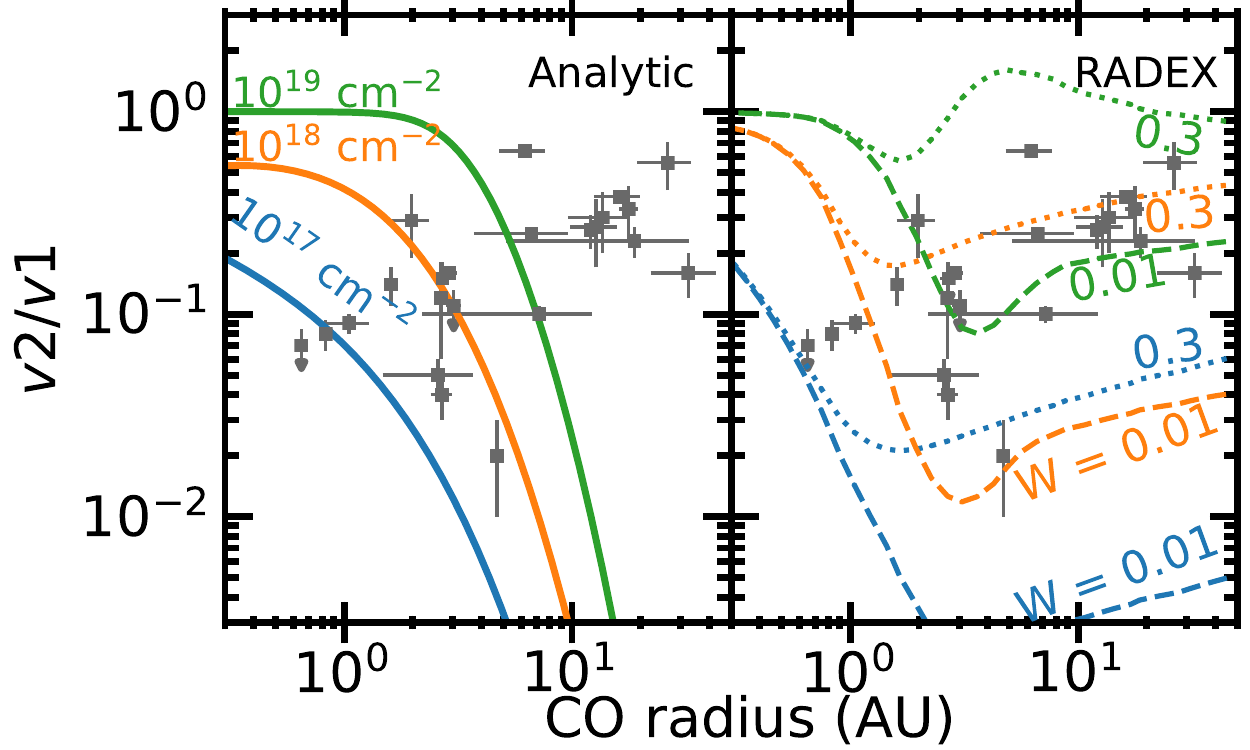}
\caption{\label{fig:ratvsrad_obs} CO vibrational ratio versus the inferred radius of emission for observational data (grey points) and analytic (\textit{left}) and RADEX (\textit{right}) model results (coloured lines). For the RADEX models, two different assumption for the radiation field are shown weakly irradiated (W = 0.01, \textit{dashed}) and strongly irradiated (W  = 0.3, \textit{dotted}) cases. For the highest column only the LTE model and the weakly irradiated RADEX model are shown. The density for the RADEX models is $10^{12}$ cm$^{-3}$. }
\end{figure}

\subsubsection{Discussion}
The vibrational ratio from these lines can be expressed as a vibrational excitation temperature. However, this only represents the excitation of the gas if both lines are optically thin. 
CO rovibrational lines get optically thick at CO columns of $10^{16}$--$10^{18}$ cm$^{-2}$ (Fig.~\ref{fig:Analytic_ratio}). Assuming a dust mass opacity of $2 \times 10^3$ cm$^{2}$ g$^{-1}$ \citep[][small grains]{Bruderer2015}, and a gas-to-dust ratio of 100 gives a \ce{H} column of $\sim 10^{22}$ cm$^{-2}$ before the dust gets optically thick at 4.7 $\mu$m. This allows for CO columns up to $10^{18}$ cm$^{-2}$ that can be detected for a canonical CO abundance of $10^{-4}$, and thus the generation of optically thick lines above the dust photosphere. Higher CO columns are possible if grains have grown beyond 1 $\mu$m or if the dust is depleted with respect to the gas in the emitting layer. Analysis of \ce{^{13}CO} lines suggest that columns of $10^{19}$ cm$^{-2}$ are not uncommon for the sources with a high vibrational ratio \citep{vanderPlas2015}. For these columns the high $v2/v1$ ratios can be explained with a temperature between 300 and 500 K. 

Equation~(\ref{eq:CO_I}) does not include the absorption of the line by dust grains nor the emission of hot dust in the region were $\tau_\mathrm{dust} < 1$. These contributions would lower the line flux, by absorbing line photons and increasing the continuum level. These effects are more pronounced at low line opacities, and so affect the $v2$ line stronger than the $v1$ lines. As such, the line ratios are overestimated.

In the case of an added dust contribution, the dust opacity sets a maximum to the CO column that can be seen, while the dust emission sets the background temperature. The CO column under consideration is thus only the CO column above the dust photosphere ($\tau_{\mathrm{dust,} 4.7 \mu\mathrm{m}} \lesssim 1$).

One critical assumption of this analysis is that both lines are formed in the same region of the disk, either under one set of conditions or under a range of conditions each of which gives rise to a similar line ratio as the region average. The idea being that if the $ v1 $ and $ v2 $ lines would be coming from different regions of the disk, this would be seen as a significantly different line shape. As the line ratios are determined on the broad component that is seen in both the $ v2 $ and $ v1 $ lines we can be sure that this assumption holds. 

\subsection{RADEX models}
Previously we have assumed a fixed excitation, parametrised by an excitation temperature. Here the excitation processes will be included explicitly by calculating the level populations from the balance between collisions, spontaneous emission and vibrational pumping. If no continuum opacity is assumed, the parameter space is four dimensional: the CO gas column, the kinetic temperature of the gas, the collision partner density (for this purpose assumed to be \ce{H2} \citep{Yang2010}\footnote{Results for H as collision partner are similar, but as the collisional rate coefficients are about an order of magnitude larger than the \ce{H2} collisional rate coefficients, the results for similar densities are shifted to more towards LTE \citep{Song2015,Walker2015}.}) and the radiation field. For the geometry, a slab that is illuminated from one side is assumed. This configuration is representative for the surface layers of proto-planetary disks, where the infrared continuum photons interacting with the gas are not along the same line of sight as the observations are taken. The pumping radiation intensity is parametrized using a 750 K black body diluted by a factor ($W$) between 0.0001 and 0.3, representative of a region at $\sim$100 times the radius of the 4.7 $\mu$m continuum emitting region and of a region very close to the continuum emitting region.

Figure~\ref{fig:RADEX_001_ratio} shows the line ratio for the $ v2$ and $ v1 $ lines from the RADEX models for different densities. This shows that for columns below $10^{17}$ cm$^{-2}$ line ratios below 0.1 are the norm. Only at high density ($> 10^{14}$ cm$^{-3}$) and high temperature ($>1300$ K) is the ratio boosted above 0.1, this is similar to the results from the analytic analysis. For $W = 0.01$, the low density results show the expected subthermal excitation of CO leading to lower line ratios compared to the LTE case. In contrast, in the $W = 0.3$ case there is a stronger contribution from excitation by infrared photons. This contribution is strongest at low temperatures where collisional excitation rates for the vibrational transitions are lowest.

\subsection{LTE vs non-LTE}
The line ratios for CO columns of $10^{17}$, $10^{18}$ and $10^{19}$ cm$^{-2}$ are plotted in Fig.~\ref{fig:ratvsrad_obs} for both the analytical and RADEX models. For these curves the temperature is assumed to scale as:
\begin{equation}
\label{eq:temp}
    T(R_\mathrm{CO}) = 1500 \left(\frac{0.4 \mathrm{AU}}{R_\mathrm{CO}}\right)^2
\end{equation}
which is approximately the dust equilibrium temperature around a star of $30\,L_\odot$. It is clear that, for these conditions the LTE models can only explain the vibrational ratios at small radii, and those only at columns $< 10^{18}$ cm$^{-2}$. 

The non-LTE RADEX models do somewhat better. With a \ce{H2} density of $10^{12}$ cm$^{-3}$, the RADEX models can reproduce the relatively low line ratios at radii smaller than 5 AU at larger columns than the analytical model. The RADEX models can also reproduce the trend in the observed data points in Figure~\ref{fig:ratvsrad_obs}, and with a strong enough radiation field, or high enough column, it can also reproduce the absolute line ratios. This indicates that high temperatures or a strongly out of LTE excitation is causing the high vibrational ratio at $R_\mathrm{CO} > 5$ AU. 

Taking into account that the infrared radiation field decreases with radius, Fig.~\ref{fig:ratvsrad_obs} implies that the \ce{CO} column responsible for the emission needs to increase with $R_\mathrm{CO}$.

\subsection{Absolute fluxes}

\subsubsection{Low vibrational ratios in the inner disk}
To further constrain the conditions of the emitting gas, it is useful to compare the absolute fluxes to the observations. First the sources with low vibrational ratios and small CO emitting radii in the lower left corner of Fig.~\ref{fig:data} are investigated. Rescaling the Herbig line fluxes from \cite{Banzatti2017} to a common distance of 150 pc leads to a range in fluxes between $4\times10^{-15}$ and $2 \times 10^{-12}$ erg s$^{-1}$ cm$^{-2}$ for the $ v1 $ line and $3 \times 10^{-16}$ and $4 \times 10^{-13}$ erg s$^{-1}$ cm$^{-2}$ for the $ v2$ line.

For these sources, the line width implies an emitting radius smaller than 5 AU. Assuming, as a conservative case, that the flux comes from the full inner 5 AU, it is possible to select condition that are able to produce both the correct $v1$ line flux and the correct line ratio. These conditions are confined between the blue and white lines in Fig.~\ref{fig:RADEX_001_ratio}. 

The low vibrational ratios and line fluxes can be reproduced with CO columns between $10^{14}$--$10^{19}$ cm$^{-2}$. More confined emitting areas would increase the lower limit of the possible columns. Temperatures between 400 and 1300 K are most likely if the CO excitation is dominated by collisions. If IR vibrational pumping dominates, higher gas temperatures are still consistent with the low vibrational ratios.

\subsubsection{High vibrational ratios at larger radii}
\label{ssc:high_vib_radex}
\begin{figure}
    \centering
    \includegraphics[width = \hsize]{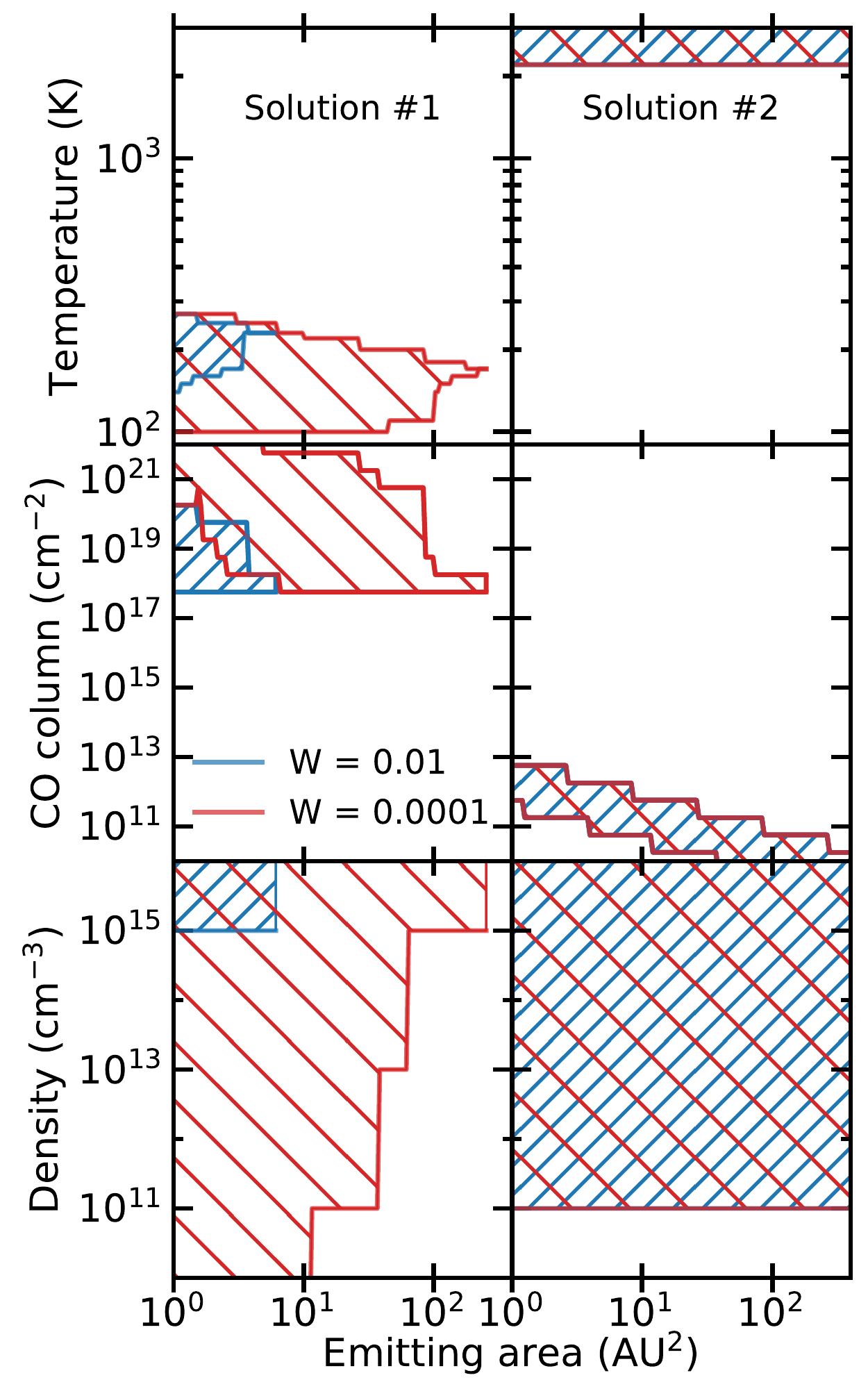}
    \caption{Parameters that can reproduce the observed CO rovibrational fluxes and line ratios for sources with $R_\mathrm{CO} > 5$ AU as function of assumed emitting area. Two solution branches are found, a low temperature (\textit{left}) and a high temperature (\textit{right}) branch. Different colours show models with different strengths of the infrared radiation field. In the second solution branch, the radiation field does not impact the solutions significantly.  }
    \label{fig:RADEX_flux_ratio_match}
\end{figure}

\begin{figure*}
    \centering
    \includegraphics[width = \hsize, page=1]{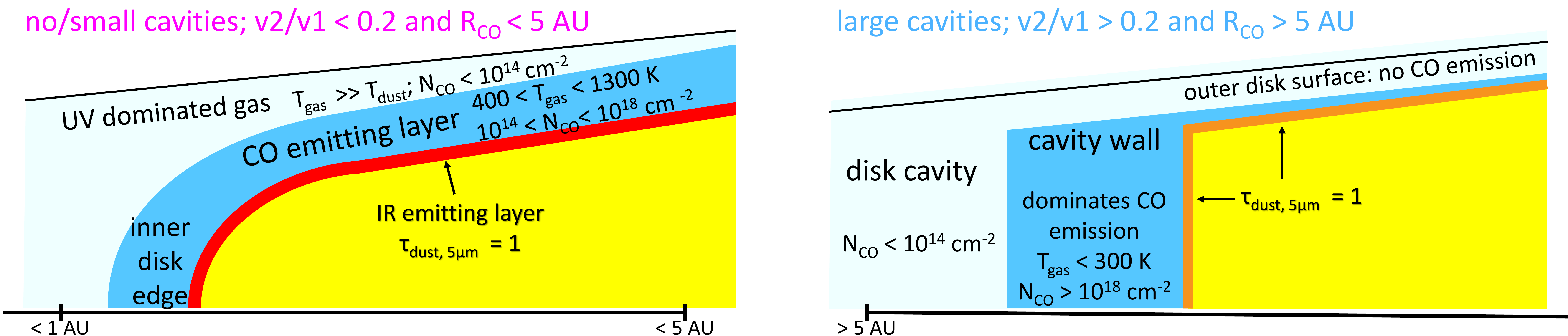}

    \caption{Summary of physical condition constraints from the RADEX models on the emitting regions of the CO rovibrational lines. Fig.~\ref{fig:Cartoon_inner} shows a version of this figure updated with the results of the full disk modelling. The constraints derived here are used to guide the DALI modelling. Regions are not shown to scale. The right panel shows solution \#1 from Fig.~\ref{fig:RADEX_flux_ratio_match} as observations of \ce{^{13}CO} ro-vibrational lines indicate the presence of large columns of \ce{CO} \citep{vanderPlas2015}.    }
    \label{fig:Schem_after_radex}
\end{figure*}

To extract the physical conditions in the emitting regions for the disks with a high vibrational ratio at large radii (low NIR group I disks), the observed fluxes and vibrational ratios were compared with the predicted fluxes from a grid of RADEX models. As the emitting area for these disks is harder to estimate than for sources with a low vibrational ratio at small radii, the emitting area was left as free parameter. As CO is coming from large radii it is expected that the near-infrared radiation field will be weak in the CO emitting area for these sources. Therefore, weaker radiation fields (W = 0.0001 and 0.01) are used in the RADEX modelling. Figure~\ref{fig:RADEX_flux_ratio_match} shows the conditions that lead to a vibrational ratio between 0.2 and 0.5 and total integrated $v1 $ fluxes between $3 \times 10^{-15}$ and $5\times 10^{-14}$ erg s$^{-1}$ cm$^{-2}$ (normalized to 150 parsec), the range of observed values for low NIR group I sources.

Within the RADEX models there are two families of solutions. For clarity these solution families have been split in Fig.~\ref{fig:RADEX_flux_ratio_match}. One solution family is characterised by low temperatures ($< 300$ K) and very high column densities ($>10^{18}$ cm$^{-2}$), the other solution has high temperatures ($> 2000$ K) and low column densities ($< 10^{14}$ cm$^{-2}$). In the low temperature family of solutions, the high line ratio comes primarily from the large columns of gas. The density is virtually unconstrained at small emitting areas and the lowest radiation fields. To allow for a large emitting area, very high densities are needed ($>10^{13}$ cm$^{-3}$), these densities are not reasonable at 10 AU, especially not in the disk surface. In the $W = 0.0001$ case, the radiation field does not dominate the excitation. The low temperature branch is also sensitive to the IR continuum radiation field. A stronger IR field moves the solutions to smaller emitting surface areas and higher densities as excitation conditions are more easily met and the line surface brightness increases. Strong IR radiation fields are not expected to be produced by local dust so a local pumping field with $W = 0.0001$ is preferred over $W = 0.01$. 

In the high temperature family the excitation of CO is dominated by the collisions with the gas. At these high temperatures both vibrational states are easily populated by collisions and the ratio in which they are populated is similar to the line ratios that is seen. As long as the density is above $10^{11}$ cm$^{-3}$, the result is density independent. Because the lines are optically thin, the line flux is given by the total amount of CO molecules giving rise to a surface area, column degeneracy. The solution is independent of the radiation field assumed. 

\subsection{Physical conditions in the CO emitting region}
\label{ssc:physical_cond_slab}

The slab models provide important constraints on the physical conditions of gas producing the observed CO rovibrational emission. In Fig.~\ref{fig:Schem_after_radex} these constraints have been put in the context of simple disk geometries. Modest temperatures ($ \lesssim 1000$ K) and columns below $10^{18}$ cm$^{-2}$ are needed to explain the low vibrational ratios at small $R_\mathrm{CO}$. These columns are most likely present in the surface layers of a dust rich inner disk and imply gas-to-dust ratios smaller than 1000, assuming that the dust is optically thick at 4.7 $\mu$m. The modest temperatures needed indicate that at these small radii, the temperature of the CO emitting gas cannot be more than a factor $\sim 2$ higher than the dust temperature, as gas that is hotter than twice the dust temperature would easily reach 1500 K, especially within the inner AU. This would create higher vibrational ratios than measured. In the next Section, the constraints from the slab models will be used as guidance for the thermo-chemical modelling, and the constrains will be updated with the results from the full disk modelling.

To explain the high vibrational ratios coming from large radii a large gap in molecular gas is needed. As these sources also have low near-infrared continuum emission and gaps have been imaged in many of them \citep[e.g.][]{Garufi2017}, a gap devoid of most of the gas and all the dust is assumed. In the case of a large dust gap, the CO column in the cavity needs to be very low, on average lower than $10^{14}$ cm$^{-2}$. If the column were higher, then the $v1$ flux from within 5 AU would be strong enough to be detected. Alternatively, it can be estimated that the surface area of optically thick CO gas within the cavity needs to be $\lesssim 0.25 $AU$^2$. 

Two families of solutions have been found from the RADEX models to fit both the line strengths and the line ratios. The first solution is shown in the left panel in Fig.~\ref{fig:Schem_after_radex} and needs low temperatures and high columns. This solution is preferred as fits of the rotational diagram of rovibrational lines of \ce{^{12}CO} and \ce{^{13}CO} for disks with a high vibrational ratio \citep{vanderPlas2015} prefer large columns $N_\mathrm{CO} \approx 10^{19}$ cm$^{-2}$ and moderate temperatures (300 -- 500 K). To be able to probe these large columns, local gas-to-dust ratios in the CO emitting regions above 100 are necessary, with many solutions needing gas-to-dust ratios of 10000.

The increase in vibrational line ratio with emitting radius seems thus to be an effect of the increase in gas-to-dust ratio of the CO emitting area, with CO coming from gas with a temperature that is coupled to the dust for both the low $v2/v1$ and the high $v2/v1$ sources. This indicates that the process that clears out the inner disk of gas in the high $v2/v1$ sources, clears out the dust as well and confines it to larger radii than the gas. This is what would be expected for a dust trap and in line with the low metallicity measured in the accreting material in these sources \citep{Kama2015}. We will discuss these scenarios in Section \ref{sec:discussion}.

\section{DALI modelling}
\label{sec:DALI}
\subsection{Model setup}

\begin{table}
\caption{\label{tab:All_mod_param} Fiducial parameters}
\begin{tabular}{l c c}
\hline
\hline
Parameter & Symbol&  Value \\
\hline
Stellar Luminosity & &30 $L_\odot$ \\
Stellar Mass & & $ 2.5 M_\odot$ \\
Effective Temperature & & $10000$ K \\
Sublimation radius & $R_\mathrm{subl}$ & 0.4 AU \\
Critical radius &$R_c$ & 50 AU \\
Disk outer radius & $R_\mathrm{out}$ & 500.0 AU \\
Gas surface density at $R_c$ & $\Sigma_c$ & 60 g cm$^{-2}$ \\
Surface density power law slope & $\gamma$ & 1 \\
Disk opening angle & $h_c $ & 0.1 \\
Disk flaring angle & $\psi$ & 0.25 \\
PAH abun. rel. to ISM & $x_\mathrm{PAH,\ ISM}$ & $10^{-20}$ \\
Large dust fraction& & 0.9 \\
Large dust settling& & 0.1 \\
Disk inner radius & $R_\mathrm{in}$ & 0.4 -- 15 AU \\
Gas-to-dust ratio & $\Delta_\mathrm{g-d}$ & 10 -- 10000 \\
\hline
\end{tabular}
\end{table}
\begin{figure}
\includegraphics[width = \hsize]{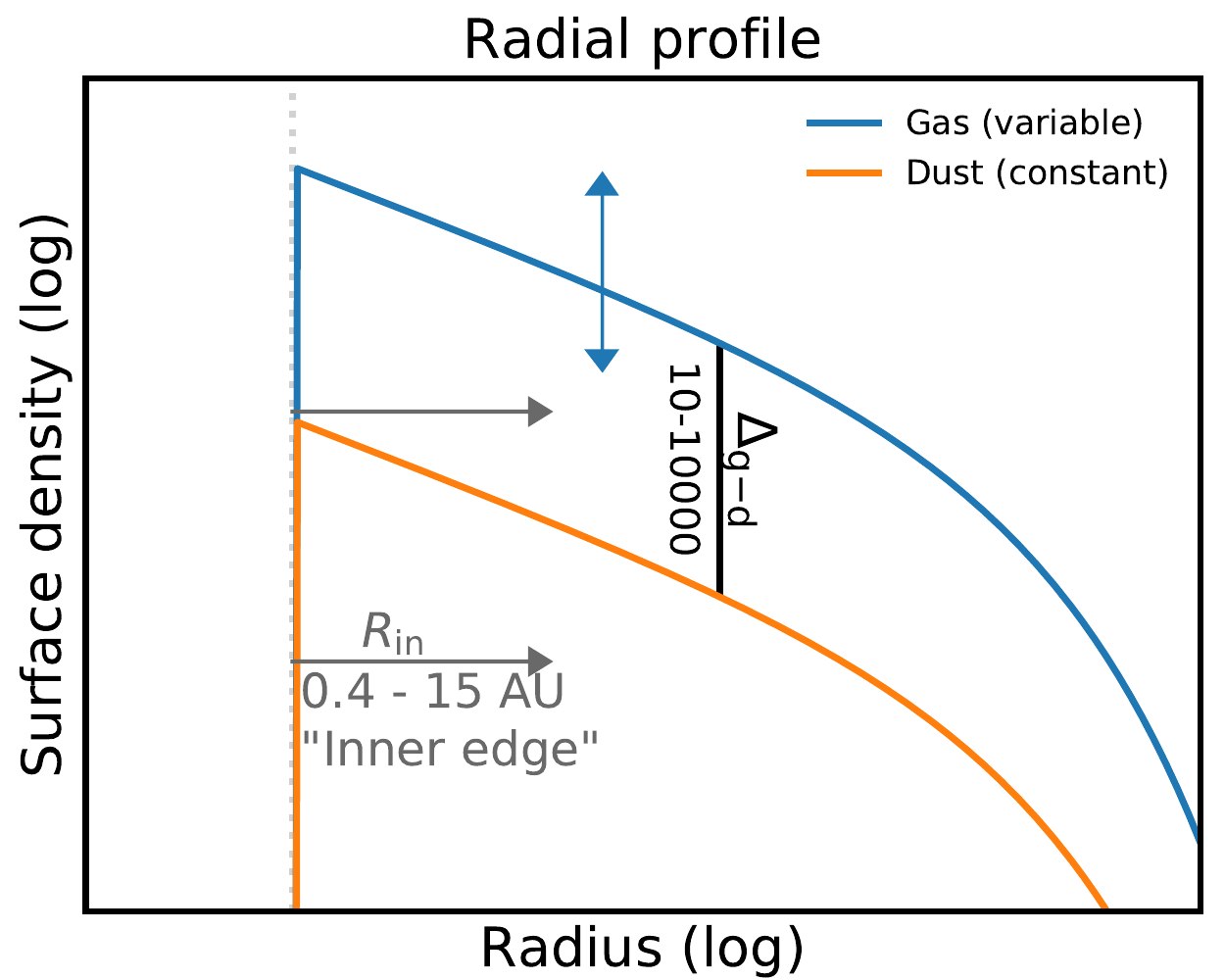}
\caption{\label{fig:Schem_mono} Schematic representation of the surface density in the monolithic models. $R_\mathrm{in}$ is the same for gas and dust and is varied between 0.4 and 15 AU, while $\Delta_\mathrm{g-d}$ is varied between 10 and 10000.}
\end{figure}
Armed with an understanding of which conditions reproduce the observations we now run a set of DALI models. Different sets of Herbig disks is modelled with the inner edge, that is, innermost radius at which gas and dust is present ($R_\mathrm{in}$), varied from the classical sublimation radius at 0.4 AU up to 15 AU (see Fig.~\ref{fig:Schem_mono}). Table~\ref{tab:All_mod_param} shows the parameters assumed for the model disks. The gas and dust surface densities are given by:
\begin{equation}
    \begin{split}
        \Sigma_{\mathrm{gas}} &= \Delta_\mathrm{g-d} \Sigma_{\mathrm{dust}} \\
        \Sigma_{\mathrm{dust}} &= \frac{\Sigma_c}{100} \left(\frac{R}{R_c}\right)^{-\gamma} \exp{\left[-\left(\frac{R}{R_c}\right)^{2-\gamma}\right]},
    \end{split}
\end{equation}
where the gas-to-dust ratio is varied between 10 and 10000. 
The thermo-chemical modelling is done with the code DALI \citep{Bruderer2012, Bruderer2013}. The standard setup is used except for a few changes that will be highlighted where relevant. Dust temperature is calculated using Monte Carlo radiative transfer. Gas temperature, chemical composition and molecular excitation are self-consistently calculated. For the thermo-chemical calculation both the \ce{CO} and \ce{H2O} molecular models have been expanded. For CO five vibrational levels, up to $v = 4$ each with 41 rotational levels, up to $J = 40$ are included, with level energies, line positions and Einstein A coefficients taken from the HITRAN database \citep{Rothman2013}. Collision rate coefficients for collisions between CO and \ce{H2} \citep{Yang2010} and \ce{H} \citep{Song2015, Walker2015} are included. The full molecule model is described in Appendix~\ref{app:CO_mol}. The molecule model for \ce{H2O} has been expanded to include vibrational lines, as these could be important for cooling in the regions that \ce{CO} is emitting. For \ce{H2O} the rovibrational datafiles from LAMDA\footnote{Leiden Atomic and Molecular DAtabase \url{http://home.strw.leidenuniv.nl/~moldata/} \citep{Schoier2005}} are used \citep{Tennyson2001, Barber2006, Faure2008}. The line profiles are extracted for the CO $ v2$ and $ v1 $ transitions using the raytracer as described in \cite{Bruderer2012}. For the ray tracing a disk inclination of 45$^{\circ}$ and distance of 150 parsec is used.

The extracted line profiles are then convolved to match a resolving power of $R = 100000$, and noise is added to achieve a similar signal-to-noise as in the observations ($\sim 200$). From these line profiles the emitting radius (from the line width) and vibrational line ratio are extracted using the same method as used for observational data by \cite{Banzatti2015}.

For some models the gas temperature and chemistry are not calculated self consistently. These are the LTE models in Fig.~\ref{fig:ratvsrad_mono} and the $T_\mathrm{gas} = T_\mathrm{dust}$ model in Appendix~\ref{app:excitationtest}. In these models the gas temperature is set equal to the dust temperature as calculated by the dust radiative transfer and the CO abundance is parametrised by: 
\begin{equation}
\label{eq:CO_param}
x_{\ce{CO}} = 10^{-4} \times \frac{A_V}{1 + A_V},
\end{equation}
with $A_V$ the visual extinction as calculated from the continuum radiative transfer. For large $A_V$ the CO abundance converges to the canonical value of $10^{-4}$, at $A_V < 1$ the CO abundance is decreased from the canonical value to mimic the effects of photo-dissociation. The CO abundance globally agrees well with the CO abundance from the thermo-chemical model. These simplified models have been run in LTE conditions and in non-LTE by explicitly calculating the excitation ($T_\mathrm{gas} = T_\mathrm{dust}$ model).

\subsection{Model results}

\subsubsection{$v1$ line flux and $v2/v1$ ratio}
\label{ssc:Mono_models}
\begin{figure}
\includegraphics[width = \hsize]{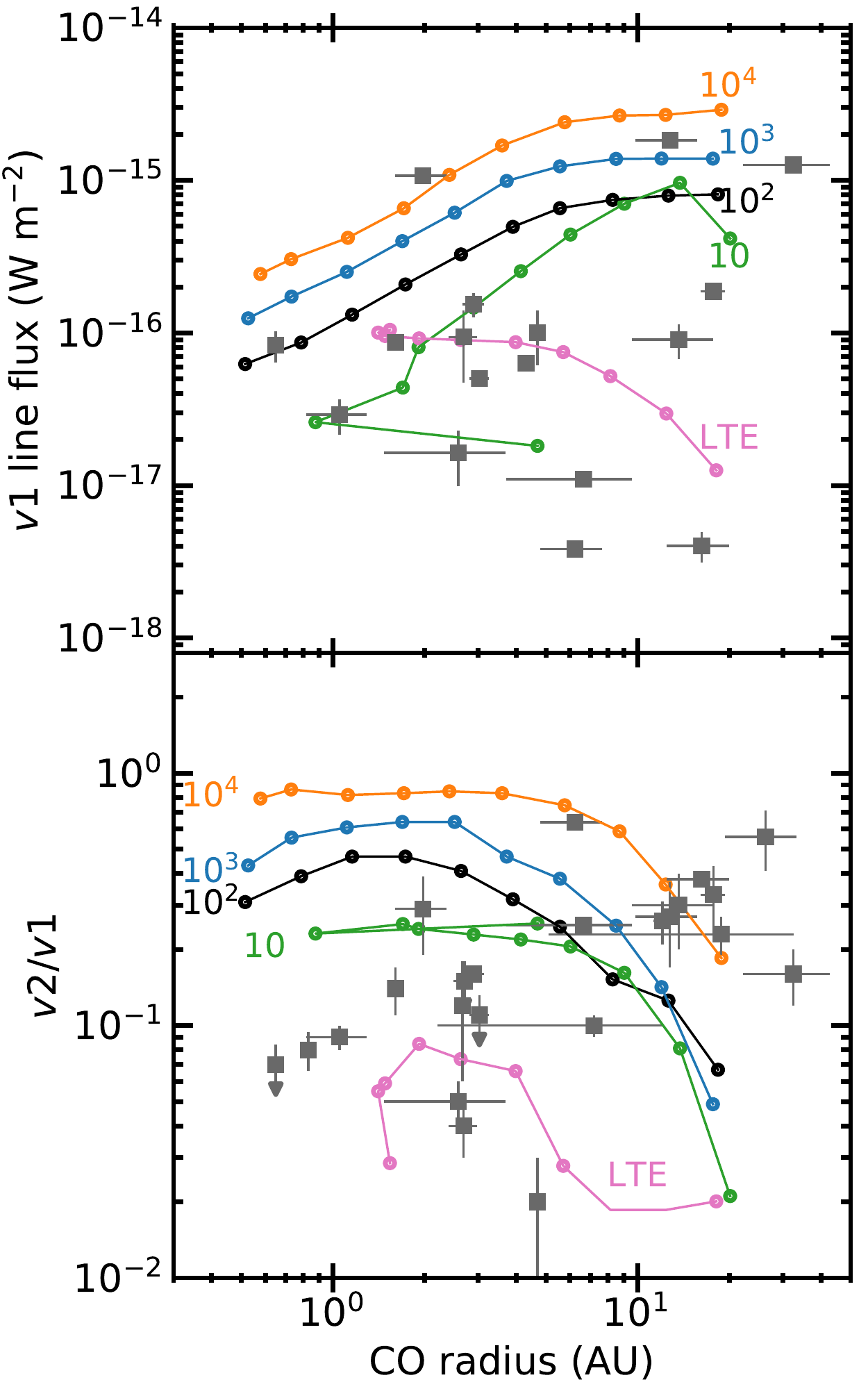}
\caption{\label{fig:ratvsrad_mono} $v1$ line flux (\textit{top}) and vibrational ratio of CO (\textit{bottom}) versus the inferred radius of emission for observational data and DALI model results. Lines connect the dots in order of inner model radius. Labels indicate the gas-to-dust ratio for the thermo-chemical models, the LTE model also has a gas-to-dust ratio of 100.} The model with the largest cavity has the largest CO radius. The dust surface density is kept constant for models of different gas-to-dust ratios. Due to missing data, not every source with a vibrational ratio in the lower panel also has a line flux in the upper panel. Clearly none of these models reproduce the trends in the data.
\end{figure}
Results of our fiducial model, with a gas-to-dust ratio of 100, are plotted in Fig.~\ref{fig:ratvsrad_mono} as the black points. The vibrational ratio from these models shows exactly the opposite trend from the data. The model vibrational ratio is roughly flat with a value around 0.4 for CO radii less than 2 AU, while at larger radii the line ratio decreases. 
The line-to-continuum ratios and line fluxes for the models are generally too high (Fig.~\ref{fig:ratvsrad_mono}, top). At small CO emitting radii line fluxes are consistent with the highest observed fluxes, at large CO emitting radii the model line fluxes are a factor $\sim$ 10 higher than the average flux.

The flux is dominated by optically thick lines coming from the inner edge of the model. The gas temperature in the emitting region is higher than $\sim 600$ K in all models. This means that the CO rovibrational lines are emitted at wavelengths longer than peak of the relevant Planck function. As a result, the line flux of these optically thick lines scales linearly with the gas temperature. The gas temperature in the emitting regions decreases slowly with increasing inner model radius and is almost constant for models with $R_\mathrm{in}$ between 1.4 and 10 AU. While the total area of the inner edge scales as $A \propto R_\mathrm{in}^{2 + \psi}$, not all of this area contributes to the emission. The emission is dominated by two rings, at the top and bottom of the inner edge wall. These rings are situated in the region where the dust temperatures are higher than the midplane dust temperature and the CO excitation is still thermalized with the gas. In the models the vertical extend of this region increases slightly with radius, leading to faster than linear growth of the emitting area. Coupled with the constant or slowly declining temperature with radius leads to a roughly linear relation between inner model radius and CO $v1$ flux.

The effect of the gas-to-dust ratio on the vibrational ratio is also shown in Fig.~\ref{fig:ratvsrad_mono} (bottom). All models show a similar trend: with increasing CO radius, the $v2$/$v1 $ drops. Models with an increased gas-to-dust ratio produce higher $v2 $/$v1 $ line ratios. This is due to the larger column of gas that can emit, leading to more optically thick lines, driving up the $v2 $ lines compared to the $v1 $ lines. Even so, the models with the lowest gas-to-dust ratio still have a $v2$/$v1$ ratio larger than most of the observed disks for emitting radii of less than $\sim 4$ AU. For models with the smallest cavities the line becomes undetectable at gas-to-dust ratios lower than 10. With increasing gas-to-dust ratio, there is also an increasing $v1$ line flux, indicating that the emitting area is getting larger.

The LTE models in Fig.~\ref{fig:ratvsrad_mono} show a very different behaviour to the thermo-chemical models. This is fully due to the LTE assumption because the parametrisation of the CO abundances and the assumption of coupled gas and dust temperature have only very small effects of the line ratios (see Appendix~\ref{app:excitationtest}). The LTE models consistently have lower vibrational ratios than the fiducial models, due to the fact that neither the IR pumping nor excitation due to self-absorption are included. Together these processes explain the different vibrational ratios between the fiducial and LTE models. The LTE models with small cavities have vibrational ratios and CO emitting radii that are consistent with with observations. For the disks with larger cavities, the LTE models cannot come close to the observations, indicating that non-LTE processes are definitely important for gas as large radii. The effect of infrared pumping and UV pumping has also been studied in Appendix~\ref{app:excitationtest} but removing IR pumping and including UV pumping only has marginal effects on the excitation and neither can explain the discrepancy between the data and the models. The LTE models consistently show line fluxes that are in good agreement with the data. 

That the LTE models seem to do so well, certainly for the low $v2/v1$ sources, is puzzling. The LTE assumption only holds for the CO rovibrational lines if the local gas density is above $\sim 10^{16}$ cm$^{-3}$. These high densities are only expected near the disk midplane and not at the disk surface. 

Beyond 5 AU there are non-LTE models that overlap with the data and there is a suggestion that higher gas-to-dust ratios are needed to explain the increase in vibrational ratio with increasing CO emitting radius. However, fluxes are high for these models, $10^{-15}$ W m$^{-2}$ at 150 parsec, about a factor of 300 higher than most observed high vibrational line ratio sources. The DALI models do not show a hot, tenuous layer of CO such as would be needed for solution \#2 of Fig.~\ref{fig:RADEX_flux_ratio_match}. They show instead that with high gas-to-dust ratios, and thus high CO columns the right vibrational ratios can be reproduced, consistent with solution \#1 of Fig.~\ref{fig:RADEX_flux_ratio_match}.

\subsubsection{Line profiles}
\label{ssc:LineProfs}
\begin{figure*}
\sidecaption
\includegraphics[width = 12cm]{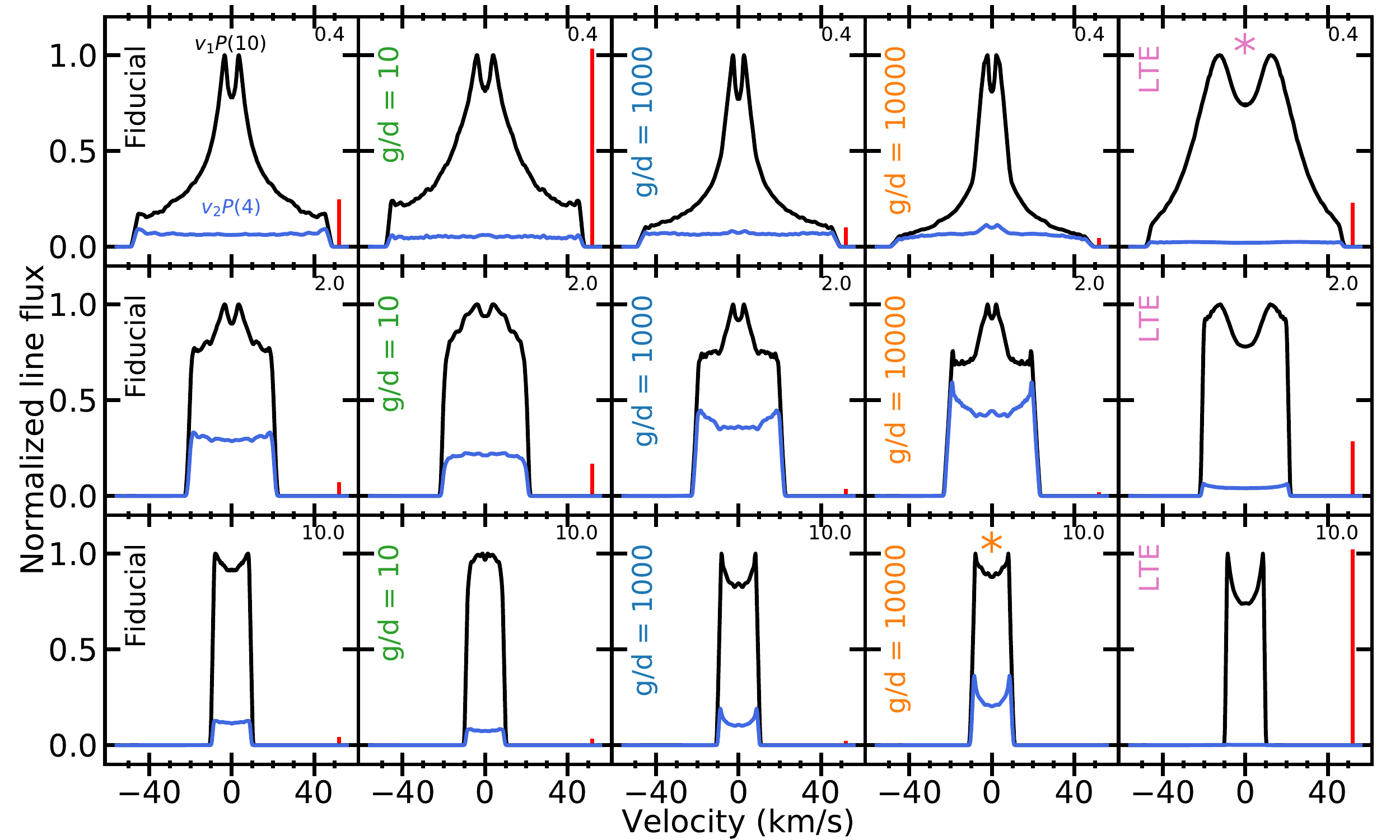}
\caption{\label{fig:lineprofiles_sub} Normalised model line profiles for the $ v1 $ (black) and the $ v2$ (blue) lines for a subset of the models at the native resolution of the model, $R = 10^6$. The text on the left of each panel denotes the model set. The top right corner of each panel denotes the inner radius of the model. The vertical bar in the bottom right of each panel shows 0.03 (top two rows) or 0.3 (bottom row) of the continuum flux density. Two models with a "*" match both $R_\mathrm{CO}$ and $v2/v1$ for a subset of the data. All lines are modelled assuming a 45 degree inclination. No noise has been added to these lines, noise-like features in the line profiles are due to the sampling of the DALI grid.}
\end{figure*}
As shown in Fig.~\ref{fig:ratvsrad_mono}, the extracted line ratios and emitting areas of most models are not able to explain the observed behaviour, especially the low line ratios in the inner disk. It is thus necessary to take a closer look at the predicted line profiles, and compare those with the observed line profiles (Fig.~\ref{fig:obslineprofiles}) for an explanation for this mismatch. 
The line profiles for subsets of DALI models are shown in Fig.~\ref{fig:lineprofiles_sub} (line profiles for all models are shown in App.~\ref{app:lineprofs}). 

The models with small holes ($< 2$ AU) show a clear difference between the full thermo-chemical models and the LTE models. The full DALI models consistently show a two component line structure. There is a broad, nearly top hat, component of the line which is present in both the $v1 $ and the $v2 $ lines, and a more strongly peaked line profile that is very weak in the $v2 $ line. This second component compares well to the line profile of HD 31648 in Fig.~\ref{fig:obslineprofiles}.  

The total line flux and the line ratio are seen to increase with increasing gas-to-dust ratio. Furthermore, the $v1 $ line profile gets narrower with increasing gas-to-dust ratio, consistent with the emitting area getting larger for higher gas-to-dust ratios.

None of the observations show the broad plateau-like feature that is in our model line profiles with small $R_\mathrm{in}$ ($< 2$ AU). This indicates that the inner rim of the model disk needs to be adapted to fit the data. Mostly, the $v2 $ flux from the inner disk wall needs to be strongly reduced. The LTE models show that low vibrational ratios are produced if the gas, dust and CO excitation are thermalised. To thermalise the CO excitation densities in the emitting area of more than $\sim 10^{16}$ cm$^{-3}$ are necessary, this is an increase in density of about 4 orders of magnitude compared to the current density of the inner disk wall. Another option would be to lower the \ce{CO} abundance from the inner rim regions by at least 4 orders of magnitude, removing most of the contribution of the inner rim to both the $v1$ and $v2$ lines.

The line profiles from models with an inner radius of 10 AU generally show a narrow double peaked profile in both lines, indicating that the directly irradiated inner edge is contributing most of the flux in both transitions. This is consistent with the very steep line profiles without low-level wings seen from disks with a high vibrational ratio (e.g. IRS 48, Fig.~\ref{fig:obslineprofiles}). The line ratio strongly depends on the gas-to-dust ratio in the disk surface: high gas-to-dust ratios lead to higher vibrational ratios as the $v1$ line opacity increases. Higher gas-to-dust ratios also lead to larger $v1$ fluxes. The LTE models with large cavities have no detectable $v2$ emission. It is, however the only model $v1$ line for which the flux is within the observed range; the non-LTE models overpredict the flux. 

Comparison of the line profiles in Fig.~\ref{fig:obslineprofiles} and Fig.~\ref{fig:lineprofiles_sub} indicates that for observed disks with low vibrational ratios, the line profiles can be well reproduced by models that have a small cavity radius except that these models have a plateau-like contribution to the line profile in the inner disk. This indicates that  emission from the disk surface agrees with the observed line profiles and line ratios. This is consistent with the analytical and RADEX analysis which predict low vibrational ratios for disk surface conditions. 

\begin{figure}
\includegraphics[width=\hsize]{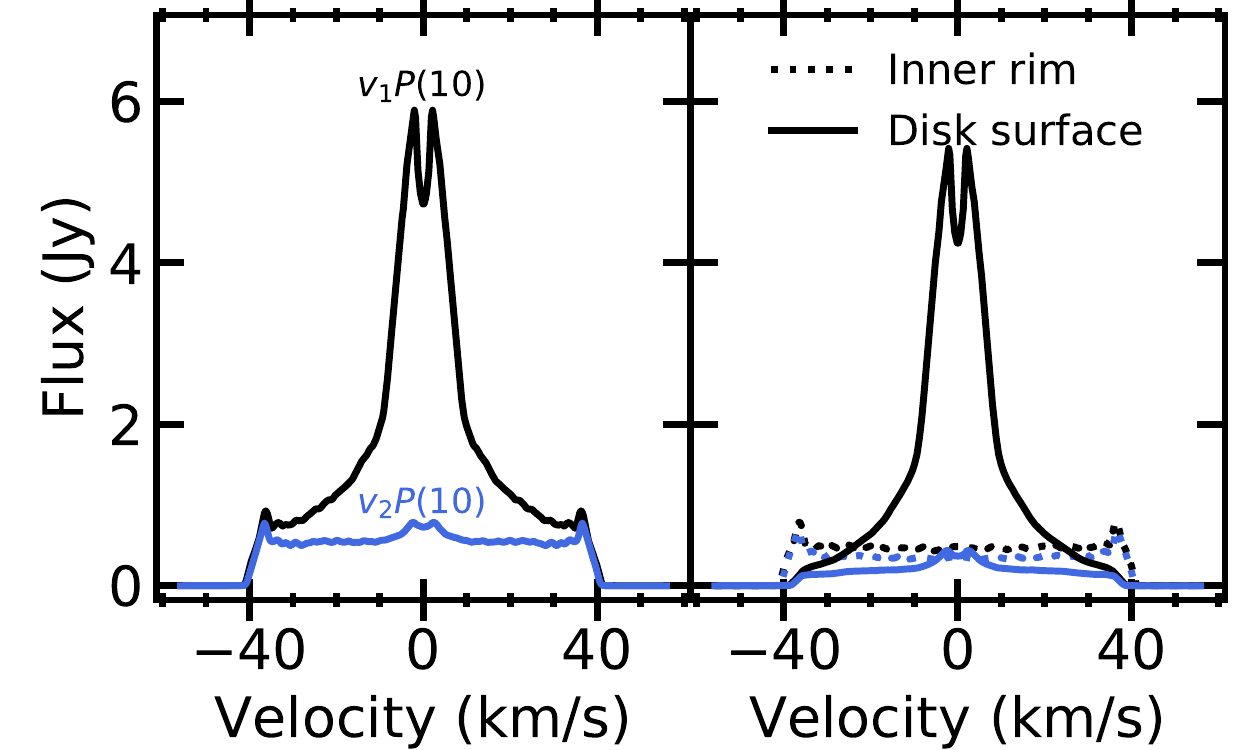}
\caption{\label{fig:sep_cont_gd10000} Line profiles for the models with an inner cavity of 0.6 AU and a gas-to-dust ratio of 10000. In the right hand plot contributions from the inner rim and disk surface are separated. }
\end{figure}

Using the spatial information in the model image cube, the emission was decomposed into a disk surface and a disk inner rim component. Fig.~\ref{fig:sep_cont_gd10000} shows the original (continuum subtracted) and decomposed line profile for the model. The line profile cleanly separates into a broad, high line ratio component coming from within 0.63 AU and a narrowly peaked, low line ratio component from the rest of the disk. This suggest that the models strongly overestimate the flux coming from the inner rim. The implications of this will be discussed in Sec.~\ref{ssc:dis_inner}. 

\subsection{Disk surface emission}
\label{sec:Sep_inner_rim}

\begin{table*}
\centering
\caption{\label{tab:subbed_params} Model variations for the models with subtracted edge contributions.}

\begin{tabular}{l c c | c | c c c | c c c | c c c}
\hline
\hline
\multicolumn{3}{c}{Inner radius variation} & & \multicolumn{3}{c}{g/d = 100}  & \multicolumn{3}{|c}{g/d = 1000}   & \multicolumn{3}{|c}{g/d = 10000}  \\
 & $R_\mathrm{in}$ (AU) & & $F_\mathrm{NIR}$ &$v2/v1$&$R_\mathrm{CO}$ (AU) & $F_{v1}$\tablefootmark{a}&  $v2/v1$&$R_\mathrm{CO}$ (AU) & $F_{v1}$\tablefootmark{a} &$v2/v1$&$R_\mathrm{CO}$ (AU)& $F_{v1}$\tablefootmark{a}\\
\hline
\# 1. &  0.4 & &0.14 & 0.03 & 2.3 & 4.6 & 0.05 & 2.4 & 9.7 & 0.33 & 3.7 & 20.9 \\
\# 2. &  0.6 & & 0.16 & 0.02 & 2.8 & 4.9 & 0.04 & 3.2 & 10.4 & 0.32 & 5.4 & 21.8 \\
\# 3. &  1.35 & & 0.18 & 0.04 & 15.2 & 5.5 & 0.03 & 20.7 & 12.1 & 0.18 & 4.5 & 23.8 \\
\hline
\multicolumn{3}{c|}{Outer radius variation} & & \multicolumn{3}{c}{g/d = 100}  & \multicolumn{3}{|c}{  }   & \multicolumn{3}{|c}{g/d = 10000}  \\
& $R_\mathrm{out}$ (AU)& & $F_\mathrm{NIR}$ &$v2/v1$&$R_\mathrm{CO}$ (AU)& $F_{v1}$\tablefootmark{a}  &  & &  &$v2/v1$&$R_\mathrm{CO}$ (AU)& $F_{v1}$\tablefootmark{a}\\
\hline
\# 1. &  3 & & 0.13 & 0.18 & 1.4 & 2.0 &     &     &     & 0.32 & 1.5 & 7.0 \\
\# 2. &  5 & & 0.13 & 0.12 & 2.5 & 2.6 &     &     &     & 0.25 & 2.3 & 9.0 \\
\# 3. &  8 & & 0.14 & 0.07 & 3.2 & 3.0 &     &     &     & 0.42 & 1.1 & 10.4 \\
\# 4. & 500 & & 0.14 & 0.03 & 2.3 & 4.6 &     &     &     & 0.33 & 3.7 & 20.9 \\
\hline
\multicolumn{3}{c|}{Flaring variation} & & \multicolumn{3}{|c}{g/d = 100}  & \multicolumn{3}{|c}{  }   & \multicolumn{3}{|c}{g/d = 10000}  \\
 & h (rad) & $\psi$ & $F_\mathrm{NIR}$ & $v2/v1$&$R_\mathrm{CO}$ (AU)& $F_{v1}$\tablefootmark{a} &   &  &  &$v2/v1$&$R_\mathrm{CO}$ (AU)& $F_{v1}$\tablefootmark{a}\\
\hline
\# 1. &  0.02 & 0.0 & 0.07 & 0.47 & 0.8 & 0.2 &     &     &     & 0.35 & 0.7 & 1.4 \\ 
\# 2. &  0.1 & 0.0 & 0.45 & 0.18 & 1.4 & 3.5 &     &     &     & 0.67 & 0.80 & 16.6 \\
\# 3. &  0.02 & 0.25 & 0.03 & 0.08 & 12.2 & 0.4 &     &     &     & 0.03 & 2.5 & 5.9 \\
\# 4. &  0.1 & 0.25 & 0.14 & 0.03 & 2.3 & 4.6 &     &     &     & 0.33 & 3.7 & 20.9 \\
\hline
\end{tabular}
\tablefoot{
\tablefoottext{a}{$v1$ line flux ($\times 10^{-14}$ erg cm$^{-2}$ s$^{-1}$)}}
\end{table*}

\begin{figure*}
\sidecaption
  \includegraphics[width=12cm]{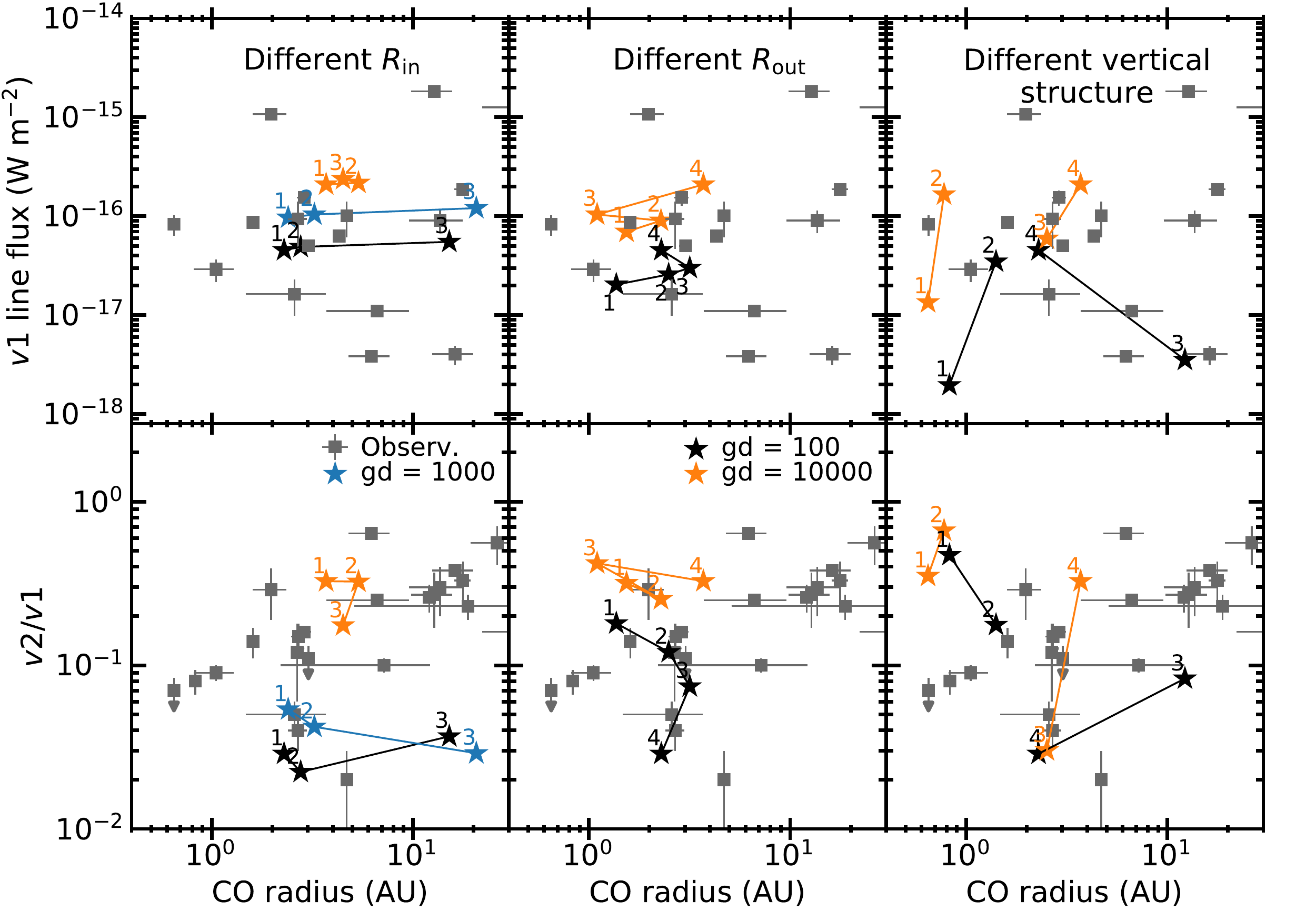}
     \caption{\label{fig:subtracted_ratios} $v1$ flux (\textit{top}) and CO vibrational ratio (\textit{bottom}) versus the inferred radius of emission for observational data and model results. The contribution from the inner edge has been subtracted from the models spectra before analysis. Models show variation in inner radius (\textit{left panel}), variation in outer radius (\textit{middle panel}) and variation in flaring and disk height (\textit{right panel}). In the left and middle panels lines connect points in increasing order of the parameter varied. In the right panels, lines connect models with the same flaring angle and thus the difference between connected points show the effect of a change in thickness of the disk. Table~\ref{tab:subbed_params} lists the parameters varied for these models. Models with a gas-to-dust ratio of 100 and 1000 are better at reproducing the vibrational ratio than the models with gas-to-dust ratios of 10000. Variation in the vibrational ratio can be reproduced by variations in the disk structure, but no single parameter explain all the variation. }
\end{figure*}

The removal of the line contribution from the inner rim makes it possible to make a direct comparison between observations and the flux from the disk surface. We restrict ourselves to model disks with a small inner cavity size (< 1.5 AU) as for these radii the vibrational ratio is most strongly over predicted in the models. The inner rim region from which the line emission is removed originally produces $\sim 40\%$ of the $v1$ flux and $\sim$90\% of the $v2$ flux. This region also accounts for $\sim$90\% of the 4.7 $\mu$m continuum flux in the model. As before, different inner disk radii and gas-to-dust ratios are studied. On top of that, for models with an inner radius of 0.4 AU and gas-to-dust ratios of 100 and 10000, the outer radius and vertical scale height and flaring are also varied. Table~\ref{tab:subbed_params} gives an overview of the varied parameters and model results. Figure~\ref{fig:subtracted_ratios} compares results of the DALI models without a contribution of the inner rim to the observed data. 

By isolating the emission from the disk surface, low vibrational ratios can be obtained at small CO radii. Increasing the gas-to-dust ratio increases the vibrational ratio and the $v1$ line flux, while only slightly increasing the CO emitting radius. Increasing the inner cavity radius to more than 1 AU causes the CO emitting radius to increase beyond 10 AU for gas-to-dust ratios of 100 and 1000. No sources with such a narrow CO line and a low vibrational ratio are seen. 

Truncating the outer disk, by removing all material beyond a radius of 8, 5 or 3 AU moves the emission inward and generally increases the vibrational ratio, because the emission has less contribution from larger radii and colder gas. The more truncated disks also have lower $v1$ fluxes, while the NIR continuum emission is not reduced compared to their full disk counterparts. As expected, a more flared disk has emission from further out, and is vibrationally colder, than a geometrically flatter disk. Lowering the scale height moves the emission further out for a non-flared disk, while for flared disks the emitting radius is reduced.

Overall, figure~\ref{fig:subtracted_ratios} shows that emission from the disk surface, especially with gas-to-dust ratios of 100 or 1000, can match the observed CO line fluxes and vibrational ratios at small radii. Different inner radii disk cannot explain the full extent of the data. Restricting the emitting region, in this case by truncating the disk, or changing the vertical structure of the inner disk helps in reproducing the spread in vibrational ratio and CO radius. This indicates that rovibrational CO emission is tracing substructures in the inner disk surface. 

Comparing the model line profiles (Fig.~\ref{fig:lineprofiles_sub}) with the observed line profiles (Fig.~\ref{fig:obslineprofiles}) reveals that there is only one disk that is matched well with a full, flared disk (HD 31648, also known as MWC 480). All other line profiles are better matched with a very flat or even truncated model. The ubiquity of emission at large radii in the models, but not in the data, implies that the inner disk structure of the observed disks is different from the smooth, flared geometry assumed in the model.

\subsection{$T_\mathrm{gas} \approx T_\mathrm{dust}$}
The removal of the inner rim for the small $R_\mathrm{in}$ models (Sec.~\ref{sec:Sep_inner_rim}) and the lower temperatures in the rounded models with large $R_\mathrm{in}$ (Appendix~\ref{app:outerdisk}) allow us to reproduce line widths, line strengths and vibrational ratios. All these models have in common that the gas and dust temperature in the emitting area are similar, with 20\% temperature differences in the surface layers of the disks with small holes and difference below 50\% for the inner walls of disks with large cavities. Conversely, models that over predicted the flux or vibrational ratio generally had gas temperatures that were at least twice as high as the dust temperature. 

These results seem contradictory with results from \citet{Bruderer2012} who modelled the pure rotational high $J$ CO lines in HD 100546, a low-NIR group I source in our sample. Bruderer et al. find that they need a gas temperature that is significantly higher than the dust temperature to explain the $v=0$ high $J$ CO rotation diagram. However, the emitting area for the high $J$ and rovibrational CO lines is not the same. The high $J$ lines come from the surface of the outer disk, while the rovibrational CO lines come from the cavity wall. This difference in emitting region is due to the difference in critical density of the transitions. The critical density of the CO rovibrational lines is around $10^{15}$ cm$^{-3}$ while the $v=0, J = 32-31$ transition has a critical density around $10^{7}$ cm$^{-3}$. The CO rovibrational lines are thus coming from denser ($\sim 10^{10}$ cm$^{-3}$), better thermalised gas than the high $J$ CO lines that can be effectively emitted from the more tenuous, thermally decoupled surface layers. 

The thermo-chemical models by \citet{Thi2013} show a slightly stronger gas-dust temperature decoupling in the inner 10 AU at the CO emitting layer with the gas temperature being 2--3 times higher than the dust temperature. The temperature of the CO emitting layer in \citet{Thi2013} is still within the 400--1300 K range. Further testing will have to be done to see if this hotter layer can also reproduce the low vibrational ratios that are observed. 

In T-Tauri disks, models of the \ce{H2O} mid-infrared observations have invoked a decoupling of gas and dust temperatures high in the disk atmosphere to explain \textit{Spitzer} observations \citep[e.g.][]{Meijerink2009}. In these models this decoupling happens at densities below $10^{9}$ cm$^{-3}$, which is lower than the density of the gas that produces most of the CO rovibrational lines of $> 10^{10}$ cm$^{-3}$. Our models also show a strong decoupling of gas and dust temperatures ($T_\mathrm{gas} > 3 \times  T_\mathrm{dust}$) in this layer, but no CO rovibrational lines are emitted from there. 

Observations of optically thin CO ro-vibrational lines, i.e. high J \ce{^{12}CO} and CO isotopologue lines, can be used to directly probe the gas temperatures predicted here. The high J \ce{^{12}CO} will most likely be more sensitive to the hotter, upper or inner layers of the disk atmosphere and so a higher gas temperature would be inferred from these lines compared to the \ce{^{13}CO} and possibly \ce{C^{18}O} ro-vibrational lines.

\begin{table*}[]
    \centering
    \caption{Summary of physical constraints from modelling results. }
    \label{tab:constraints}
    \begin{tabular}{l | c c c}
        \hline 
        \hline
        
        CO emission  & Conditions & Model & Comments \\ 
        \hline
         \multirow{3}{*}{} & $T < 1500$ K, $N_\mathrm{CO} < 10^{18}$ cm$^{-2}$ & Slab LTE & Fig.~\ref{fig:Analytic_ratio} \\
           & $T = 400 - 1300$ K, $10^{14} <N_\mathrm{CO} < 10^{18}$ cm$^{-2}$ &  RADEX & Fig.~\ref{fig:RADEX_001_ratio}\\
        $v2/v1 < 0.2$ and    & No CO in dust free gas & Slab LTE, RADEX & Sec.~\ref{ssc:physical_cond_slab} \\
        $R_\mathrm{CO} <  5$   &  g/d$ \lesssim 1000$ & DALI & Fig.~\ref{fig:subtracted_ratios} \\
                        & No CO at the inner rim & DALI & Sec.~\ref{sec:Sep_inner_rim} and Fig.~\ref{fig:subtracted_ratios}\\
                        & Radially constrained emitting area & DALI & Sec.~\ref{sec:Sep_inner_rim} and Figs.~\ref{fig:obslineprofiles},~\ref{fig:lineprofiles}~and~\ref{fig:lineprofiles_sub}\\
        \hline
        & $T \lesssim 300 $K, $N_\mathrm{CO} > 10^{18}$ cm$^{-2}$ \textbf{or}& \multirow{2}{*}{Slab LTE, RADEX} & \multirow{2}{*}{Figs.~\ref{fig:Analytic_ratio},~\ref{fig:RADEX_001_ratio}~and~\ref{fig:RADEX_flux_ratio_match}}  \\
        $v2/v1 > 0.2$ and   & $T \gtrsim 1000 $K, $N_\mathrm{CO} < 10^{14}$ cm$^{-2}$&  &  \\
        $R_\mathrm{CO} >  5$ & g/d$ > 10000$ & DALI & Figs.~\ref{fig:ratvsrad_mono}~and~\ref{fig:lowflux_highvib}\\
         & Rounded edge, $T_\mathrm{gas}< 300$ K, $N_\mathrm{CO} > 10^{20}$ cm$^{-2}$ & DALI & Sec.~\ref{app:outerdisk} \\
        \hline
    \end{tabular}

\end{table*}

\begin{figure*}
    \centering
    \includegraphics[width = \hsize]{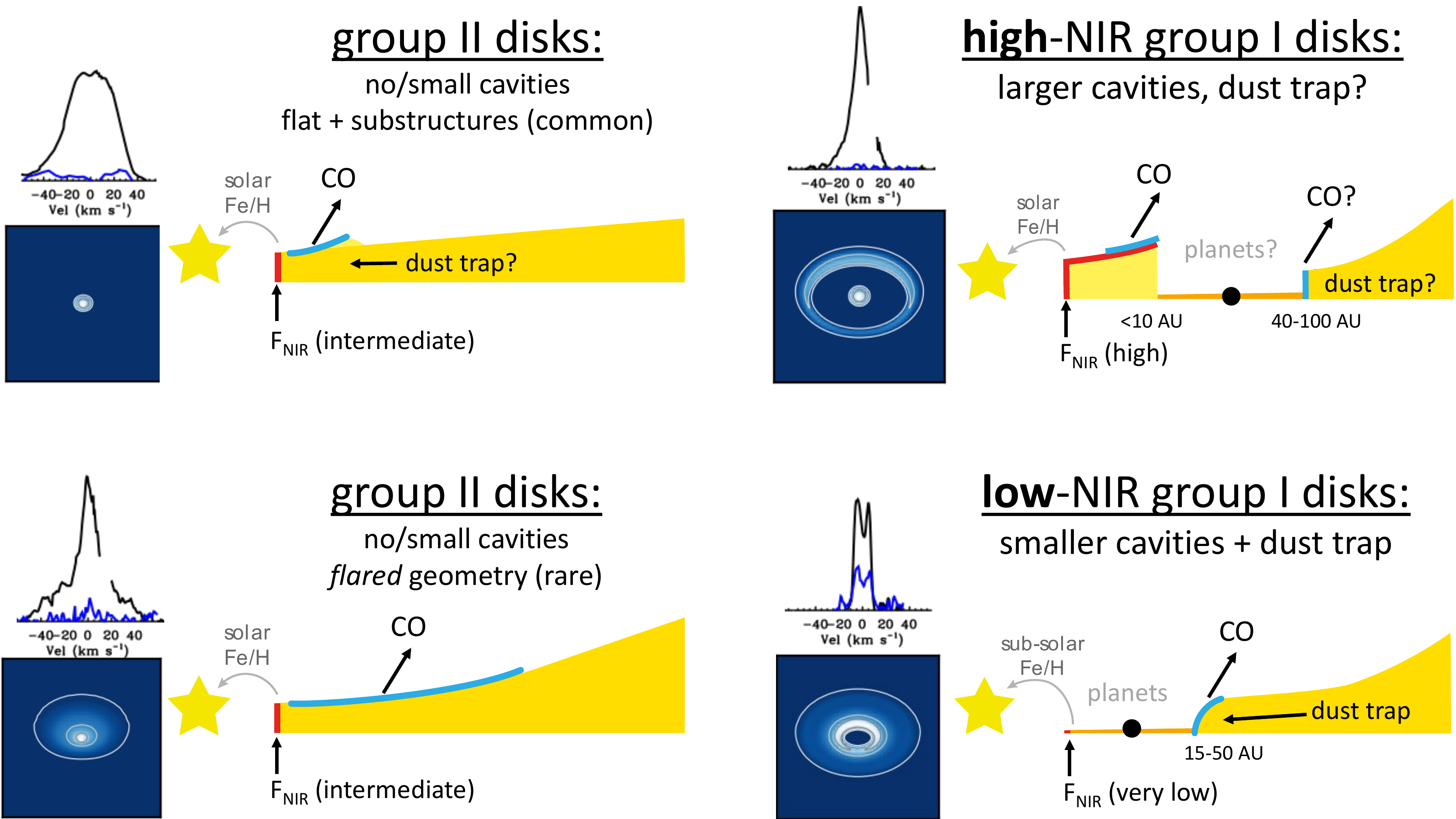}
    \caption{\label{fig:Cartoonend} 
    Typical line profiles, simulated images and inferred disk proposed disk structures for four types of disks identified in the Herbig sample. Near-infrared continuum and CO emitting areas are shown in red and blue respectively. The simulated images show the velocity integrated CO $v1$ line flux. These images are discussed in more detail in Sec.~\ref{ssc:futureobs}. The disk structures are updated versions of those shown in Fig.~\ref{fig:data}.}
\end{figure*}

\section{Discussion}
\label{sec:discussion}

Our modelling results show that we can reproduce the observed CO emission with low vibrational ratios at small radii versus high vibrational ratios at large radii (Fig.~\ref{fig:data}) under different and separate conditions. Low vibrational ratios measured at small disk radii require a CO column below $10^{18}$ cm$^{-2}$ and a temperature between 400--1300 K. These conditions naturally occur in the denser $>10^9$ cm$^{-3}$ surface layers of the disk. Emission from a dust-free inner region, and from an inner disk rim directly irradiated from the star, are ruled out based on the high $v2/v1$ that would be produced under these conditions which are not observed. Line velocity profiles indicate that most group II disks have an emitting area that is radially narrower than what a flared disk model produces. Thus, flared group II disks should be rare, in agreement with the currently accepted paradigm.

High vibrational ratios measured at large disk radii require instead an inner disk region strongly devoid in CO ($N_{\ce{CO}} < 10^{14}$ cm$^{-2}$), i.e. the region that would otherwise produce the high-velocity CO line wings that are not observed in these spectra. In the emitting region at larger radii, where CO is still present, the gas must be cold ($< 300$ K) and CO columns must be high ($> 10^{20}$ cm $^{-2}$). To provide these conditions, a high gas-to-dust ratio is necessary ($> 10000$) coupled with a density structure that allows for efficient cooling. Models with midplane number density and column density that increase with radius are able to match the line flux, vibrational ratio and $R_\mathrm{CO}$. The large columns and low gas temperatures are consistent with \ce{^{13}CO} observations \citep{vanderPlas2015}. These constraints on the disk physical structure, and their relative models, are summarized in Table.~\ref{tab:constraints}.  

Figure~\ref{fig:Cartoonend} shows four representative CO line profiles, simulated images, and cartoons of the disk structures proposed to produce the observed emission as based on the combination of this and previous analyses. In Sec.~\ref{ssc:dis_inner} and Sec.~\ref{ssc:dis_outer} we will link these structures to the physical and chemical processes proposed to produce them.

Three different disk structures for the low vibrational ratios at small radii are shown in Figure~\ref{fig:Cartoonend}. Two of these apply to group II disks. They are divided into structures with a compact (upper left) and extended (lower left) rovibrational CO emitting area. CO line profiles and infrared excess show that abundant gas and dust is present within $\sim 5$~AU, so any existing disk cavities must be smaller or the dust extends to the sublimation radius. The stellar abundances for these sources are consistent with solar Fe abundances, so transport of dust to the star is relatively unhindered \citep{Kama2015}. The NIR and CO flux are not emitted from the same region: the IR flux mostly comes from the inner dust edge that is directly irradiated by the star, while the CO must only be emitted from the disk surface. The major difference between the two group II structures is that those with compact CO emission must have a non-flared geometry confining the CO emitting region, possibly with inner disk substructures shadowing the rest of the disk, further confining the emission. The one group II disk with extended CO emission (HD 31648) needs to have a more flared geometry or have a slow molecular disk wind, analogous to those seen in T-Tauris \citep{Pontoppidan2011, Brown2013}. Based on this sample we conclude that both molecular disk winds, as well as flared group II disks should, be very rare. In the case of a flared disk, the size of the emission is measurable is by spectro-astrometry on 8-meter class or IFU spectroscopy on 30-meter class telescopes. 

The third structure giving rise to low vibrational ratios at small radii is that of the high-NIR group I disks (upper right). These disks have large cavities (typically the largest found in Herbigs, see Table \ref{tab:Av_vals}), but they have a residual inner dust disk/belt that produces the high near-infrared flux. These disks therefore have a gap between inner and outer disk. CO emission comes from the inner disk, again from the disk surface in order to produce the very low vibrational ratios measured in the data. The high near-infrared flux instead, higher than the group II disks, must have come from a larger emitting area than in the group II disks, possibly from both the inner edge and surface of the inner disk. The solar Fe abundance in the surface layers of the star is an independent indicator of accretion from a still gas- and dust-rich inner disk \citep{Kama2015}, possibly implying efficient filtration of small dust from the outer disk to the inner disk.

The high vibrational ratios at large radii can all be explained with a single structure (bottom right in Figure~\ref{fig:Cartoonend}). These disks have an inner cavity that is strongly reduced in CO surface density. These inner cavities seem to be on average smaller in size than those imaged in high-NIR group I disks (see Table \ref{tab:Av_vals}). The rovibrational CO emission comes from the cavity wall, i.e. the inner edge of the outer disk, which must be rich in molecular gas but strongly depleted in dust. This structure combined with the low metallicity of the material that is accreted on the stellar surface \citep{Kama2015}, indicates that the dust is efficiently trapped at some radii larger than $R_\mathrm{CO}$ for these disks. The most appropriate explanation for these deep gas cavities and dust traps currently seems to be that they are caused by giant planets and not by photo-evaporation or dead-zones \citep[see also][]{vanderMarel2016}.

\subsection{Implications for sources with low $v2/v1$ at small radii}
\label{ssc:dis_inner}
\subsubsection{No CO at the inner rim}

\begin{figure*}
    \centering

    \includegraphics[width = \hsize, page = 2]{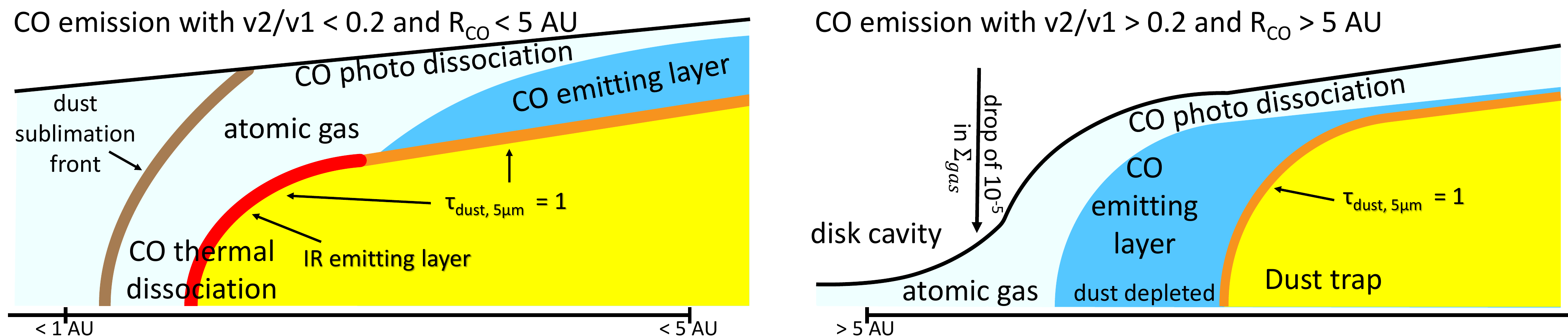}

    \caption{Sketches of the upper right quartile of a disk cross section of the preferred configuration of the CO emitting region in the case of low $R_\mathrm{CO}$ and low $v2/v1$ (group II and group I high NIR, \textit{left}) and large $R_\mathrm{CO}$ and high $v2/v1$ (group I low NIR, \textit{right}). This figure is an update to Fig.~\ref{fig:Schem_after_radex} including the DALI results. Relevant radial scales for the inner and outer edge are shown on the bottom left and right corner of each sketch.
    }

    \label{fig:Cartoon_inner}
\end{figure*}

The good match between the line profiles of disks without a contribution from the inner disk edge (Sec.~\ref{sec:Sep_inner_rim}) indicates that CO is not present within or around the dust sublimation radius in any of these disks. The left plot in Figure~\ref{fig:Cartoon_inner} shows the proposed structure of the inner disk of a group II source as inferred from the data. The dust disk in this case reaches to the dust sublimation radius, forming an inner dust rim. Our modelling suggests that, for Herbig disks, the gas gets heated to high enough temperatures ($>3000$ K) near the sublimation radius to keep the gas atomic (see App.~\ref{app:thermdiss} for a discussion on the chemistry), at least until the gas is hidden under the dust infrared photosphere. This is not seen in the thermo-chemical models, however, the gas and dust temperature in this region are very uncertain. Dust sublimation impacts both the dust temperature structure, as well as the gas temperature by changing the composition of the gas.

At larger radii the gas can cool enough to become molecular above the dust infrared photosphere. The expectation is that the phase change from atomic to molecular gas also induces a strong extra cooling effect, lowering the gas temperature to the dust temperature. This seems to be the most appropriate scenario to produce the low vibrational ratios measured at such inner disk radii. 

This scenario assumes that the dust disk extends all the way to the dust sublimation radius. However, even if there is a small inner hole in the dust disk, the radiation field should still be strong enough to heat the gas to temperatures above 3000 K and keep the gas atomic at the inner edge of the dust cavity, at least within the small inner cavities that have been explored above in Fig.\ref{fig:subtracted_ratios}. 

The CO abundance structure proposed in our modelling, with the low columns and temperatures and the absence of molecular gas within the inner dust rim, also naturally produces very weak or no CO overtone ($\Delta\nu = 2$) emission. This is consistent with a large survey of 2 $\mu$m CO emission towards Herbig AeBe stars, with detection rates as low as $7\%$ \citep{Ilee2014}. 

The dissociation of \ce{H2} and \ce{CO} as proposed here, should leave a large ($N_{\ce{H}} \gtrsim 10^{18}$ cm$^{-2}$), hot ($T> 3000$ K) atomic or ionised reservoir around the dust inner radius. Velocity or spatially resolved atomic lines, such as can now be measured with near-IR interferometry \citep[e.g.][ and with VLTI-GRAVITY, \citet{Gravity2017}]{Eisner2014, GarciaLopez2015}, can thus be used to test if \ce{CO} and \ce{H2} are indeed being dissociated around the inner edge of the disk.

\subsubsection{Group II disks have more mass within 5 AU than group~I high-NIR sources}
Both the group II disks and the high-NIR group I disks show low vibrational ratios at small radii, with the latter showing larger radii and lower vibrational ratios on average (Table \ref{tab:Av_vals}). At the same time these group I disks have a higher NIR excess than the group II disks: this is thought to be due to a vertically more extended dust structure in the inner disk \citep[see e.g. discussions in][]{Maaskant2013,Banzatti2018}.
A vertically more extended structure will lead to lower gas densities in the surface layers of the disk. A large population of small grains is needed to populate the tenuous surface layers and convert stellar flux into the observed bright NIR flux. These conditions naturally lead to larger $R_\mathrm{CO}$ and lower vibrational ratios.
In fact, low densities slow down the chemical formation of CO and observable abundances of CO are thus only produced at lower UV fluxes, further from the star. Furthermore, a larger population of small grains has a higher NIR opacity per unit mass of gas, so the visible column of CO is smaller than for group II disks. A lower gas density also helps in lowering the excitation in the $v =2$ state (Fig.~\ref{fig:RADEX_001_ratio}), thereby lowering the vibrational ratio and providing a good explanation for the measured difference from the group II disks.

Source specific modelling of the rovibrational lines of CO and its isotopes, fitting the full rovibrational sequence using non-LTE models can be used to further constrain the density in the inner regions of the these disks. Furthermore if the grains are indeed small, and the dust mass in the inner disk is low, the continuum might be optically thin at ALMA wavelengths, allowing for inner disk mass and grain size measurements. 

The radial extent of CO emission in the high-NIR group I objects is harder to estimate than for the group II objects as the observed CO lines are intrinsically narrower. Spectro-astrometric measurements of HD 142527 indicate that the CO emission extends up to $\sim 5$ AU, twice the measured emitting radius \citep{Pontoppidan2011}, and nicely matching an inner dust belt detected by \citet{Avenhaus2017}. A narrow emitting region would imply that CO only emits in the inner dust disk and is absent in the disk gap. None of the high NIR group I sources show the double peak structure expected from such a narrow emitting ring. This could be due to emission from the outer disk cavity wall and surface filling up the centre of the observed spectral line, but higher spectral and spatial resolution observations are needed to confirm this scenario. 

\subsubsection{Inner disk CO emission is confined by substructures}

\begin{figure}
    \centering
    \includegraphics[width = \hsize]{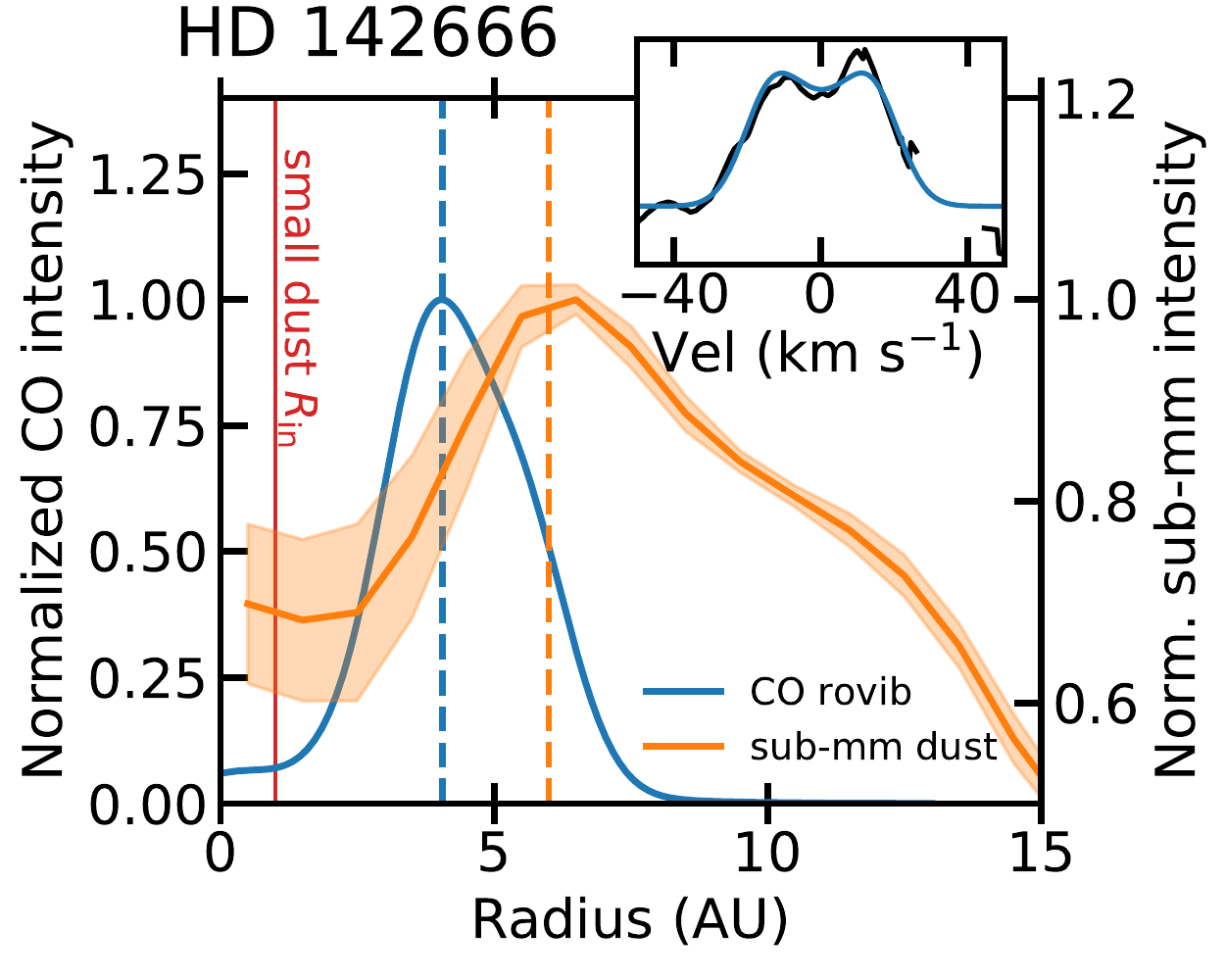}
    \caption{Radial intensity cuts for the sub-millimeter dust from \citet{Huang2018} and the radial intensity as inferred from the CO rovibrational line profile of HD 142666. Vertical dashed lines show the maximum of the CO and sub-millimeter dust intensity. The CO emission is clearly contained within the bright sub-millimeter ring at 6 AU. The inset on the top right shows the observed line profile (black) and the fitted profile (blue). The vertical red line shows the inner edge of the dust disk at $\sim1$~AU as inferred from IR interferometry \citep{Schegerer2013}.}
    \label{fig:HD142666_comp}
\end{figure}

One disk in this sample provides exceptional insight into the inner dust and gas structures: HD 142666. Its $R_\mathrm{CO}$ of $\sim3$~AU is relatively large for a group II object, and a bright ring at $\sim6$~AU has recently been found in sub-millimeter dust continuum images taken at very high angular resolution with ALMA \citep{Andrews2018, Huang2018}. Figure~\ref{fig:HD142666_comp} shows the comparison of the inner part of the sub-millimeter radial continuum intensity profile with the CO radial emission profile. The CO radial emission profile was derived by fitting a flat Keplerian disk intensity model to the observed line profile. The observed line profile and the flat disk fit can be seen in the inset in Fig.~\ref{fig:HD142666_comp}. 
The CO rovibrational emission is confined within the inner edge of the sub-millimeter dust ring, indicating that the process that is producing this sub-millimeter ring also confines the CO rovibrational emission. This could happen if the bright sub-millimeter ring traces a vertically extended dust structure that shadows the disk beyond, preventing CO emission from larger radii. 

Our fit to the CO line profile for HD 142666 shows that the inner 1 AU of the disk is devoid of emission (Figure~\ref{fig:HD142666_comp}). This would be consistent with IR interferometric observations that report an inner disk radius of 1 AU \citep{Schegerer2013}, significantly larger than the dust sublimation radius expected at 0.3 AU as based on the stellar luminosity. This indicates that a small cavity has formed in this disk; similar cavities could also be present in the other group II disks that have $R_\mathrm{CO} \gtrsim 2 R_\mathrm{subl}$. These small cavities should still allow for efficient transport of both gas and dust (in a $\sim$ 100--1 ratio) from the inner disk edge to the star, as the accretion rates are normal and the stellar abundances for these sources close to solar \citep{Kama2015,Banzatti2018}. The lower sub-mm intensity implies a lower dust surface density at 1 AU. If we assume there is no continuous build up of material around the sub-millimeter ring, then the lower surface density in the inner disk gap should be accompanied with an increase in the velocity of the accretion flow. This would be consistent with an inner dead-zone edge, possibly due to the thermal ionisation of alkali metals \citep{Umebayashi1988}. Another option would be the presence of a giant planet within the small cavity with a saturated or leaky dust trap. In either case there would be a dust trap or traffic jam. This raises the tantalizing possibility that all group II disks with $R_\mathrm{CO} \gtrsim 1$ AU could have a bright sub-millimeter ring in the inner regions as found in HD 142666 and HD 163296 \citep{Huang2018, Isella2018}. 

\subsection{Implications for high $v2/v1$ at large radii}
\label{ssc:dis_outer}
\subsubsection{High gas-to-dust ratios by dust trapping}

The large CO columns needed to produce a high vibrational ratio at large radii indicate that around $R_\mathrm{CO}$ the gas surface density does not deviate strongly from what would be expected in absence of an inner cavity. To get the $v2$ lines bright enough, CO columns larger than $10^{20}$ cm$^{-2}$ and thus total $\ce{H}$ columns larger than $10^{24}$ cm$^{-2}$ are needed. This necessitates a drop in gas surface density that is less than two orders of magnitude at $R_\mathrm{CO}$ if CO is at the canonical abundance of $10^{-4}$ alternatively, a CO abundance of more than $\sim10^{-6}$ is needed if the gas surface density is continuous within the dust cavity. The large CO column also implies that dust is under abundant by at least two orders of magnitude at $R_\mathrm{CO}$. The high gas-to-dust ratios necessary are likely not due to strong settling as ALMA images of millimeter dust show cavities that are consistently larger than $R_\mathrm{CO}$ indicating that there is a indeed a radial segregation \citep[e.g. HD 100546, HD 97048, IRS 48, HD169142; ][]{vanderMarel2016, vanderPlas2017, Fedele2017,Pinilla2018}. This is consistent with ALMA gas observations of transition disks, showing that also in the sub-millimeter, \ce{CO} cavities are smaller than dust cavities \citep{vanderMarel2013, vanderMarel2016}.

The steep line profiles and the lack of high velocity CO emission indicates that the CO column within $R_\mathrm{CO}$ is at least six orders of magnitude lower than the column at $R_\mathrm{CO}$ for the observed disks. As CO is hard to photodissociate, a drop in the total gas surface density in the cavity is necessary. A total gas surface density drop of 5 orders of magnitude between the dust ring and the CO poor cavity is needed to produce CO columns below $10^{14}$ cm$^{-2}$ \citep{Bruderer2013}. Observations of atomic oxygen or carbon could be used to measure gas depletion factors directly, constraining the depth of the cavity and thus the mass of a possible cavity-forming planet. Fig.~\ref{fig:Cartoon_inner}, right, illustrates the disk structure as reconstructed from modelling the CO rovibrational lines in low-NIR group I disks. 

The combination of a large cavity in both gas and dust and of a radial segregation between gas and dust at the disk cavity wall fits well with what is expected for a giant planet carving an inner disk hole. A sufficiently massive planet can explain the gas depletion in the cavity together with the different cavity sizes in dust and gas, as a gas pressure maximum is created that traps dust at a slightly larger radius. Both photo-evaporation and dead zone models, instead, have problems creating transition disks that have an inner region rich in gas, but depleted in micron-sized grains \citep{Pinilla2012, Gorti2015, vanderMarel2015, vanderMarel2016, Pinilla2016}. 

\subsubsection{Not all dust traps are equal}
In Sec.~\ref{ssc:dis_inner} we compared the high NIR group I disks to the group II disks on the basis of inner disk structures. However, at long wavelengths and larger radii these high NIR disks appear more similar to the rest of the group I sources showing cavities in sub-millimeter and scattered light imaging \citep[e.g.][]{Garufi2017}. One notable difference is that the inner disk in the high NIR group I sample seems to be misaligned with respect to the outer disk \citep[e.g.][]{Benisty2017,Banzatti2018}.
Due to this misalignment, perhaps the inner disk cannot shadow the cavity wall in the outer disk and thus CO emission in high NIR group I sources could in principle include a component similar to what observed in low NIR group I disks. If so, there should be a narrow emission component with large $v2/v1$ at the center of the line. This could be the origin of the narrower $v2$ emission line observed in HD 135344B (Fig.~\ref{fig:obslineprofiles}), but even at the current high spectral resolution it is not possible to distinguish a narrower central peak in the $v1$ line. Part of the problem may also be due to flux filtration by the narrow slit (0.2'' for VLT-CRIRES), where the signal from the outer disk will be diluted by any slit that includes a bright inner disk but excludes part of the outer disk \citep{HeinBertelsen2014}. 

Another possibility is that the outer disk in the high NIR group I systems may not have as high a gas-to-dust ratio as the rest of the group I systems. A normal gas-to-dust ratio would quench the $v2$ emission, with $v1$ emission from the gap outer edge filling in the line center and producing a narrow single peak. A close to ISM gas-to-dust ratio would imply that dust is not efficiently trapped at the gap outer edge. This could fit with a scenario in which  the dusty inner disk and near solar abundances in the stellar atmosphere are replenished by dust from the outer disk. Mapping of the CO emission, either with multiple slit positions with VLT/CRIRES+ or observations with ELT/METIS integral field unit can determine the brightness and nature of the CO emission from the outer disk in high NIR group I sources (see Sec.~\ref{ssc:futureobs}). 

\begin{figure*}
    \centering
    \includegraphics[width = \hsize]{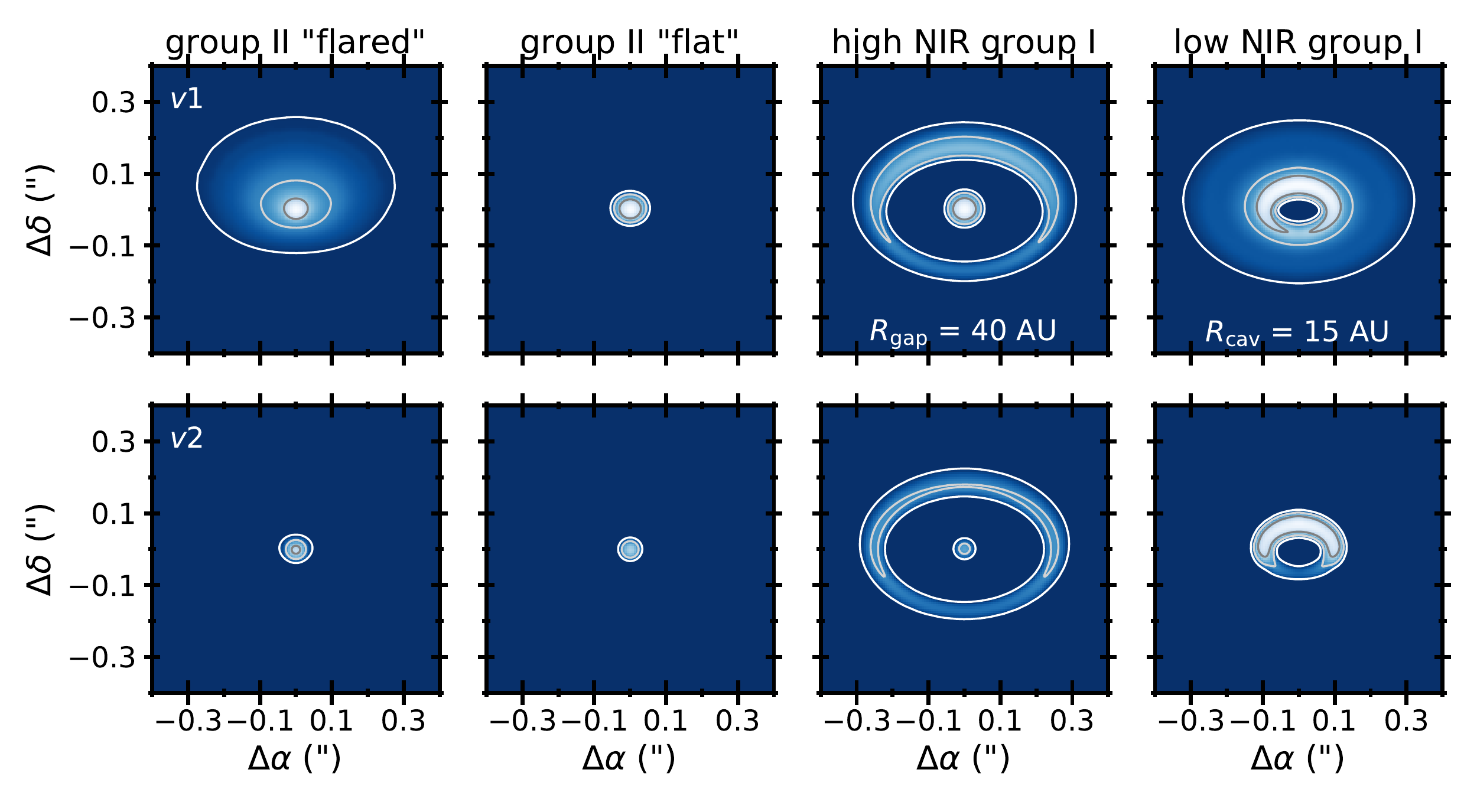}
    \caption{Simulated velocity integrated $v1 P(10)$ (top) and $v2 P(4)$ (bottom) line maps convolved to METIS resolution \citep{Brandl2014}. The colour scale is log-stretched between 0.1\% and 100 \% of the maximum of the $v1$ line flux. 
    The continuum has been subtracted before velocity integration.The contours show 0.1\%, 1\% and 10\% of the peak surface brightness. The disk geometries refer to those presented in Fig.~\ref{fig:Cartoonend}. The distance is assumed to be 150 parsec and the inclination is 45 degrees, the far side of the disk is in the north. For the "flat" geometry, a truncated gas disk (5 AU outer radius) is used. The high NIR group I image is composed by combining two models, a truncated disk model, with an inclination of 30 degrees and a disk with a 40 AU hole and a strong dust trap (so strong $v2$ emission) with an inclination of 45 degrees. No interactions between the inner and outer disk have been taken into account. }
    \label{fig:Metis_log}
\end{figure*}

\subsection{Comparison to T-Tauri disks: distribution of UV flux matters} 
Under several aspects, disks around Herbig AeBe stars are analogous to T-Tauri disks. However, in CO rovibrational emission they exhibit a very different behaviour \citep{Banzatti2015,Banzatti2018}. While the T-Tauri disks have a decreasing $v2/v1$ with increasing $R_\mathrm{CO}$, the Herbig disks show the opposite trend. This implies a significant difference in the distribution of molecular gas between Herbig and T-Tauri inner disks. This is also seen in the line profiles, since many of the T-Tauri disks have a two component CO profile: if both components originate from the Keplerian disk, it means that the CO emitting region is more radially extended than in the Herbig disks. 
This can be explained if the T-Tauri disks have CO emission from the inner rim (and from within the inner rim), while in Herbig disks CO is dissociated by high temperatures in these regions as explained above.

Both Herbigs and highly accreting T-Tauris have strong UV fields. The energy distribution as function of wavelength is very different, however. The UV field of Herbig stars is dominated by the continuum coming from the stellar surface. For T-Tauri stars, on the other hand, most of the UV comes from the accretion shocks and is emitted in emission lines, especially Lyman-$\alpha$ \citep[e.g.][]{France2014}. Both CO and \ce{H2} cannot be photo-dissociated by Lyman-$\alpha$ photons. As such the photo-dissociation of CO and \ce{H2} is much more efficient around Herbig stars. If hydrogen is mainly in atomic form, formation of other molecules such as \ce{CO2} and \ce{H2O} is significantly slowed down as molecules both need the \ce{OH} radical for their formation. This radical forms from \ce{H2 + O -> OH + H} and can be destroyed by \ce{OH + H -> H2 + O}. Only with abundant \ce{H2} can enough \ce{OH} be produced and can \ce{OH} survive long enough to form \ce{CO2} and \ce{H2O}. This could explain the lack of \ce{H2O} and \ce{CO2} emission towards Herbig AeBe disks in comparison to T-Tauri disks \citep[e.g.][]{Pontoppidan2010,Banzatti2017}.

The higher broad-band optical-UV flux in Herbig systems can have a larger impact on the gas temperature compared to the line dominated T-Tauri spectrum as more power can be absorbed by atomic and molecular electronic transitions before the dust absorbs the radiation. Finally, a larger fraction of the stellar flux can generate photo-electrons upon absorption by the dust \citep{Spaans1994}. All these effects will heat the gas more in Herbig than in T-Tauri disks, increasing CO dissociation in the former.

\subsection{Predictions for future observations}
\label{ssc:futureobs}

ELT-METIS will be a generation I instrument on the ELT \citep{Brandl2014}. It will be able to do diffraction limited imaging and IFU spectroscopy at 3--5 $\mu$m. The 39 meter mirror allows for a spatial resolution of $\sim$ 0.03'' at 4.7 $\mu$m in a 0.5'' by 1'' field of view. IFU spectroscopy will be possible at a resolving power of $R = 100000$. With these capabilities, ELT-METIS will be able to resolve CO rovibrational emission both spatially and in velocity in nearby Herbig disks. Fig.~\ref{fig:Metis_log} shows continuum subtracted, velocity integrated maps of the $v1 P(10)$ line. The four different disk structures proposed in Fig.~\ref{fig:Cartoonend} can be clearly distinguished in these images. Spatially resolving the emission will enable to study asymmetries in the spatial distribution of CO and will help in explaining the single peaked nature of CO lines observed in the high-NIR group I disks.

The CO rovibrational ratio is a good tracer of large cavities in Herbig disks, with high ratios ($>0.2$) only coming from low-NIR group I disks, intermediate ratios (0.05 -- 0.2) coming from group II disks and very low ratios ($< 0.05$) coming from the high-NIR group I disks. This could be exploited in more distant and more massive star forming regions, for instance by observing multiple Herbigs within the field of view of the \textit{JWST}-NIRSPEC multi object spectrograph, providing an efficient classification of large numbers of Herbig sources, either for more detailed follow-up or for population studies. 

In the shorter term, ground based, high sensitivity observations can be used to constrain densities in the inner disk. In all disk models, CO emission is not in LTE. This results in strongly decoupled vibrational and rotational excitation temperatures. On top of this, most of the $v1$ lines are also optically thick so there should be a break in the $v1$ rotational diagram. This is the point where the lines become optically thin. The  position and sharpness of the break critically depends on the density of the emitting area. A source by source modelling of the CO rovibrational lines over a large number of $J$ levels ($ J > 30$) can derive this density and from that the mass in the inner few AU can be constrained. The same should apply to T-Tauri disks, but as the CO line flux can have contribution from within the sublimation radius it will not be straightforward to measure the mass in the dust rich inner disk.

\section{Conclusions}
\label{sec:conclusion}

The goal of this work has been to find the physical conditions that can reproduce trends observed in inner disks of Herbig stars, in terms of the CO vibrational ratio $v2/v1$, the radius of CO emission (from the HWHM of the lines) and the NIR excess (Fig.~\ref{fig:data}).
We have studied the excitation and line profiles of CO rovibrational emission from disks around Herbig Ae stars using LTE and non-LTE slab models, as well as using the thermo-chemical model DALI. Our findings are collected in Figs.~\ref{fig:Cartoonend}~and~\ref{fig:Cartoon_inner} and our conclusions can be summarised as follows: 

\begin{itemize}
   \item \textit{Emission from the inner disk surface}: CO emission with $v2/v1 < 0.2$ at $R_\mathrm{CO} < 5$ AU is reproduced by conditions found in the inner disk surface. CO columns must be $\lesssim 10^{18}$ cm$^{-2}$ and gas-to-dust ratios $< 1000$. Gas and dust temperatures must be coupled and between 400 and 1300 K. Emission from and within the inner disk rim is ruled out on basis of the measured low vibrational ratios. A scenario in which the gas around the dust sublimation radius is hot, $> 3000$ K, is preferred to explain the absence of CO at the inner rim. At these temperatures, reactions between CO and atomic H should produce a primarily atomic gas that could be observed by IR interferometry. 
    
    \item \textit{Emission from the cavity wall}: CO emission with $v2/v1 > 0.2$ at $R_\mathrm{CO} > 5$ AU is reproduced by conditions found in a cavity wall at large disk radii. CO columns must be $> 10^{18}$ cm$^{-2}$ and gas-to-dust ratios $> 10000$. Gas and dust temperatures must be coupled and below 300 K, indicating efficient cooling of the gas. Within $R_\mathrm{CO}$ the gas surface density drops by at least 5 orders of magnitude. A high gas surface density, rather than UV pumping, is the most likely reason for the bright $v2$ lines providing high $v2/v1$ ratios. 
    
    \item \textit{Substructures in inner disks}: The broad, flat topped or double peaked line profiles that most group II sources exhibit cannot be explained by a smooth, flared disk. The radial extent of the CO emission in these sources is restricted. Flat disk models work better in matching these line profiles, but they generally still have too much flux at large radii. The outermost radius that emits in these sources thus most likely traces some variation in vertical scale-height. This could be the case for HD 142666, where all the CO emission arises from within the first resolved sub-millimeter dust ring at 6 AU.
    
    \item \textit{Dust trapping}: Small cavities in gas and dust are possible in the group II objects. The low vibrational ratios observed indicate that dust-free and molecular gas rich cavities are not present. If there are small cavities formed by planets in the sample, then they apparently do not create a very efficient dust trap. 
    This is in contrast with the low NIR group I disks that have high vibrational ratios. The large gas surface density drop and the dust poor gas necessary in these disks fit very well with predictions of giant planets producing a cavity and a strong dust trap. High NIR group I disks, instead, may have dust traps that allow for dust filtration from the outer to the inner disk, sustaining normal elemental accretion onto the star.

    \item \textit{CO as a tracer of disk cavities and inner dust belts/disks}: rovibrational CO emission can be used to identify dust cavities also in absence of direct imaging, especially when disks are further away than 200~pc; the difference in CO lines as observed in high- and low-NIR group I disks moreover shows that, in disks with large cavities, CO emission is a good tracer for residual inner dust belts/disks that may be unseen in direct imaging.
    
    \item \textit{Molecular gas within the sublimation radius}: The lack of broad, high vibrational ratio, CO emission in many of the observed sources puts strong constraints on the amounts of dust free molecular gas within the sublimation radius: $N_\mathrm{CO} < 10^{18}$ cm$^{-2}$. This upper limit is consistent with the non-detection of the CO overtone (v = 2-0) emission towards most Herbig AeBe disks. 

    \item \textit{Future METIS observations}: ELT-METIS will be able to resolve the emitting area of the CO rovibrational lines. These observations can further constrain the emitting area in group II disks. For low NIR group I disks METIS should find CO rovibrational rings within the scattered light and sub-millimeter dust cavities, while for high NIR group I disks the extent of molecular gas in the inner disk as well as the gas-to-dust ratio, and thus efficiency of dust trapping, in the outer disk can be measured. 
    
\end{itemize}

\begin{acknowledgements}

We thank Antonio Garufi for helpful discussions on imaging of Herbig disks, Inga Kamp and Daniel Harsono for help with the CO collisional rate coefficients and Paul Molliere for providing the results to the chemical equilibrium calculations. 
Astrochemistry in Leiden is supported by the Netherlands Research School for
Astronomy (NOVA).

This work is partly based on observations obtained with iSHELL under program 2016B049 at the Infrared Telescope Facility, which is operated by the University of Hawaii under contract NNH14CK55B with the National Aeronautics and Space Administration. This work is partly based on observations made with CRIRES on ESO telescopes at the Paranal Observatory under programs 179.C-0151, 093.C-0432, 079.C-0349, 081.C-0833, 091.C-0671. This work is partly based on observations obtained with NIRSPEC at the W. M. Keck Observatory, which is operated as a scientific partnership among the California Institute of Technology, the University of California, and the National Aeronautics and Space Administration. The observatory was made possible by the generous financial support of the W. M. Keck Foundation.

This project has made use of the SciPy stack \citep{Jones2001}, including NumPy
\citep{Oliphant2006} and Matplotlib \citep{Hunter2007}.
\end{acknowledgements}

\bibliographystyle{aa}
\bibliography{Lit_list}
\appendix

\section{CO molecule model}
\label{app:CO_mol}

\subsection{Rovibrational}
The CO molecule model used in this work contains 205 energy levels distributed over five vibrational states ($J = 0-40, v = 0-5$). Energy levels and line strengths have been extracted from the HITRAN database \cite{Rothman2013}. Collisional excitation by \ce{H2} and \ce{H} was included. For the CO-\ce{H2} collisional de-excitation rates the collisional rates from \cite{Bruderer2012}, based on \cite{Yang2010} have been used. The collisional rate matrix was expanded using the formalisms in \cite{Chandra2001}. 

For the H-CO collisions, the pure rotational rate coefficients were taken from \cite{Walker2015} while the rovibrational rate coefficients were taken from \cite{Song2015}. The rates from \cite{Walker2015} were used for all $\Delta v = 0$ transitions. 

\subsection{Electronic}

\label{app:CO_mol_UV}
The electronic excitation of CO has been done in post processing. From the model output the total rate into and out of each vibrational state due to collisions and photon absorption and emission was calculated. To these vibrational band-to-band rates the electronic rates are added and a new vibrational equilibrium is calculated. Using the rotational distribution from the model, a new rovibrational level distribution is calculated. This is then used for ray tracing. 

The electronic transitions and Einstein A coefficients were taken from \cite{Beegle1999}. The Einstein A coefficients were used to calculate a transition probability matrix for an absorption and subsequent emission of a UV photon. It is assumed that electronic relaxation into a vibrational level of the ground state with $v'' > 4$ will further cascade into $v'' = 4$ and will be treated as if the relaxation directly goes into $v'' = 4$. By construction the rotational distribution of the molecules was assumed to be unaffected by the electronic excitation.

\section{Excitation tests}
\label{app:excitationtest}

\begin{figure}
    \centering
    \includegraphics[width = \hsize]{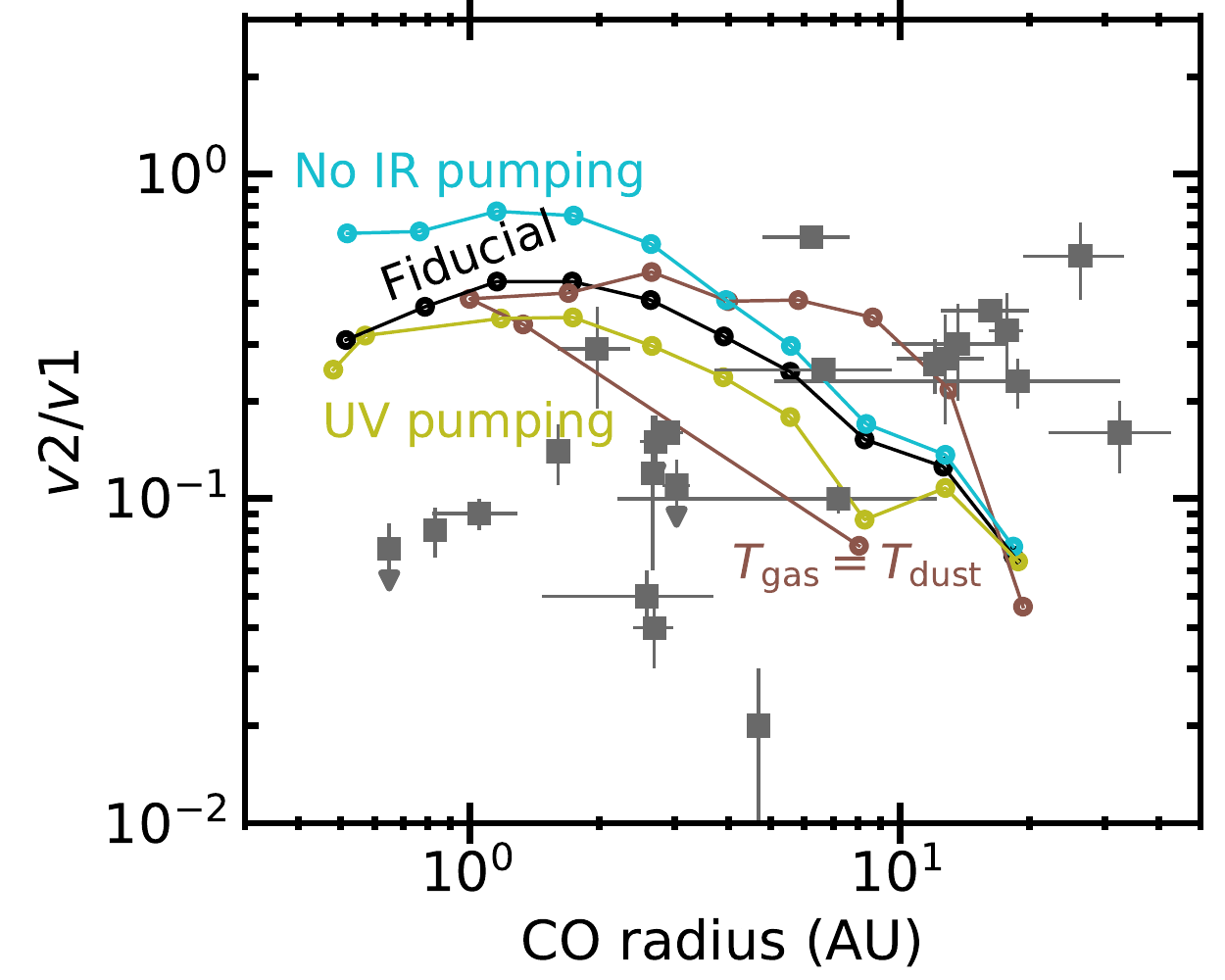}
    \caption{\label{fig:ratvsrad_mono_app} Ratio of the flux from the $v=2$ and $v=1$ levels of CO versus the inferred radius of emission for observational data and DALI model results. Line connect the dots in order of inner model radius. Labels indicate the different assumptions for the excitation calculation. All models have a gas-to-dust ratio of 100.} The model with the largest cavity is always at the largest CO radius. Different assumptions on the excitation are tested.
\end{figure}

To increase our understand of the CO line formation processes in DALI, a few variations on the standard temperature and excitation calculations have been done. In one set of models, the thermo-chemistry has been skipped and $T_\mathrm{gas} = T_\mathrm{dust}$ has been assumed. The CO abundance in these models is parameterized according to Eq.~(\ref{eq:CO_param}). The excitation has been tested without IR pumping and with UV pumping. Results have been plotted in Fig.~\ref{fig:ratvsrad_mono_app}.

Assuming equal gas and dust temperature lowers the vibrational ratio somewhat for the models with small inner radii and has slightly increased vibrational ratios for models with 5 to 10 AU gaps. For models with gaps smaller than 5 AU the vibrational ratio is still higher than the data. The models with equal gas and dust temperature also predict very low (> 0.05) vibrational ratios for models with gaps larger than 10 AU. Fig.~\ref{fig:ratvsrad_mono_app} shows that the thermo-chemical balance does influence the line fluxes and line flux ratios for CO.   

A returning topic of discussion for vibrational excitation of molecules is the treatment and relative importance of pumping by radiation of different wavelengths. In DALI the radiation that pumps the lines, radiation that is absorbed by the molecules increasing their excitation, is always assumed to come from either radially inward, or vertically upward, whichever has the lowest line optical depth. This could, depending on the situation, both over- and under-predict the effective pumping flux. For CO there are two possible ways to excite vibrational levels through the absorption of a photon. The first is the absorption of an infrared photon through the rovibrational lines around 4.7 or 2.3 $\mu$m raising the vibrational excitation by one or two quanta directly. In Fig.~\ref{fig:ratvsrad_mono_app} a comparison is made between models with and without infrared pumping, but otherwise using the same abundance and temperature structure. The general trend is that the infrared pumping lowers the vibrational ratio, especially in the inner regions of the disk. Infrared pumping increases the excitation in the $v = 1$ state more strongly than the $v=2$ state. This greatly increases the vertical and radial extent of the emitting region of the $v1$ lines increasing line flux. Although the $v=2$ state can be directly pumped from the ground state, the Einstein $A$ coefficient for these lines is small, contributing little to the overall excitation of the second vibrationally excited state. 

The inclusions of UV pumping has a small effect of the line fluxes. However it mostly effects the disks with cavities smaller than 10 AU and in these models the vibrational ratio is lowered. The regions were the UV field is strongest is in the cavity wall near the star. These regions show emission with a vibrational ratio close to unity. UV pumping and the vibrational cascade would create a vibrational ratio of $\sim 0.6$ thus lowering the contribution of the $v2$ line in the inner region in favour of the $v1$ line. At radii larger than 10 AU the UV field is so diluted that it can no longer affect the vibrational ratio. Fig.~\ref{fig:ratvsrad_mono_app} shows UV pumping without taking into account the self-absorption of UV photons by \ce{CO} molecules. Including self-absorption only changes the vibrational ratios by a few percent. 

\section{Line profiles}
\label{app:lineprofs}
\begin{figure*}
\includegraphics[width = 0.97\hsize]{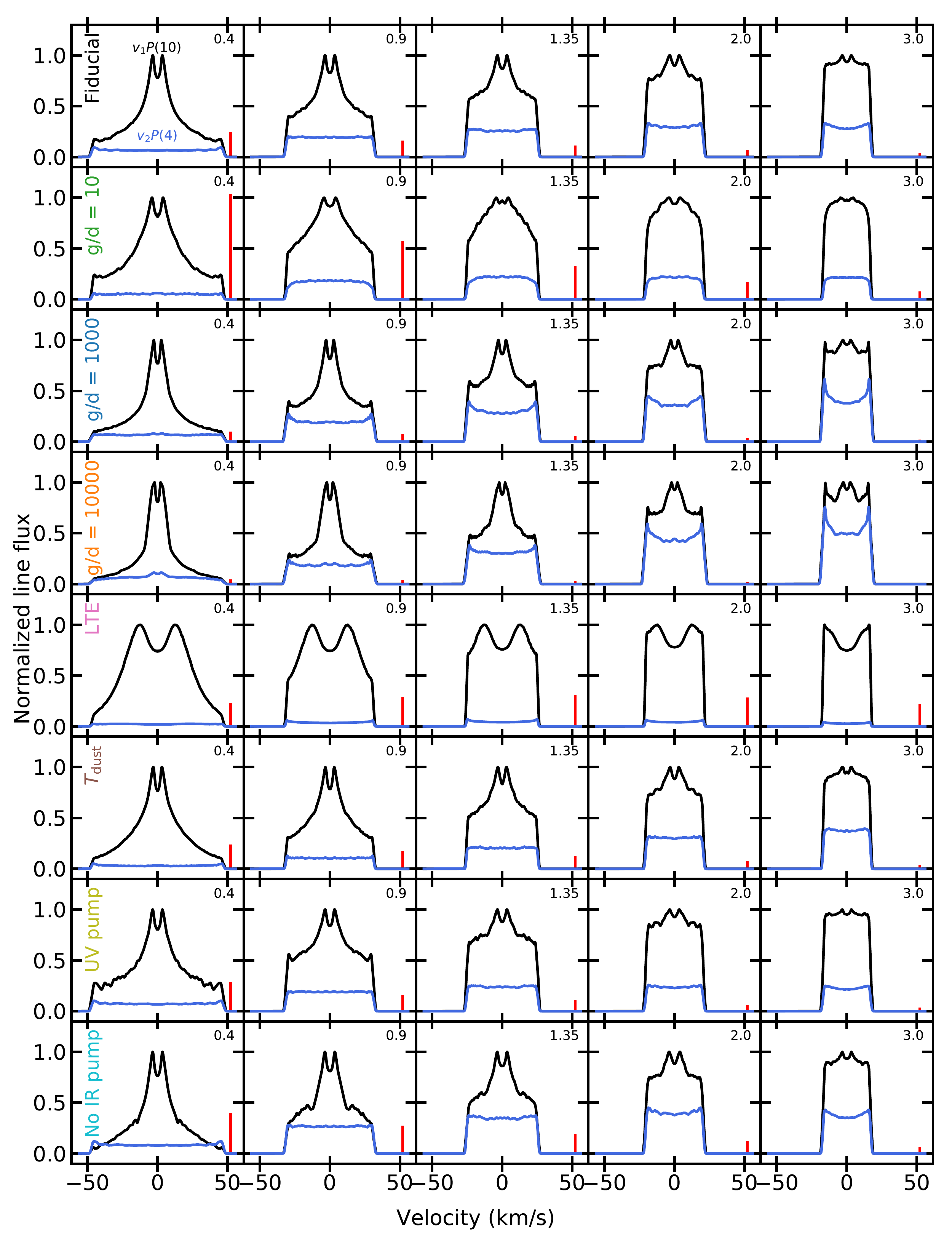}
\caption{\label{fig:lineprofiles} Normalised model line profiles for the $ v1 $ (black) and the $ v2$ (blue) lines for a subset of the models at the native resolution of the model, $R = 10^6$. The text on the left of each panel denotes the model set. The top right corner of each panel denotes the inner radius of the model. The vertical bar in the bottom right of each panel shows 0.03 of the continuum flux density. All lines are modelled assuming a 45 degree inclination. No noise has been added to these lines, features in the line profiles are due to the sampling of the DALI grid. }
\end{figure*}

Figure~\ref{fig:lineprofiles} shows the line profiles for the disks with the smaller inner holes. Most line profiles show two components, a broad component that is present in both the $ v2$ and the $ v1 $ line and a narrow component, that is mostly seen in the $ v1 $ line. The velocity of the broad component is the component used in the flux ratio extraction and radius determination. This component is consistent with emission dominated by the inner wall of the model. The narrow component is more extended and thus has to come from the disk surface. 

\begin{figure*}
\includegraphics[width = 0.83\hsize]{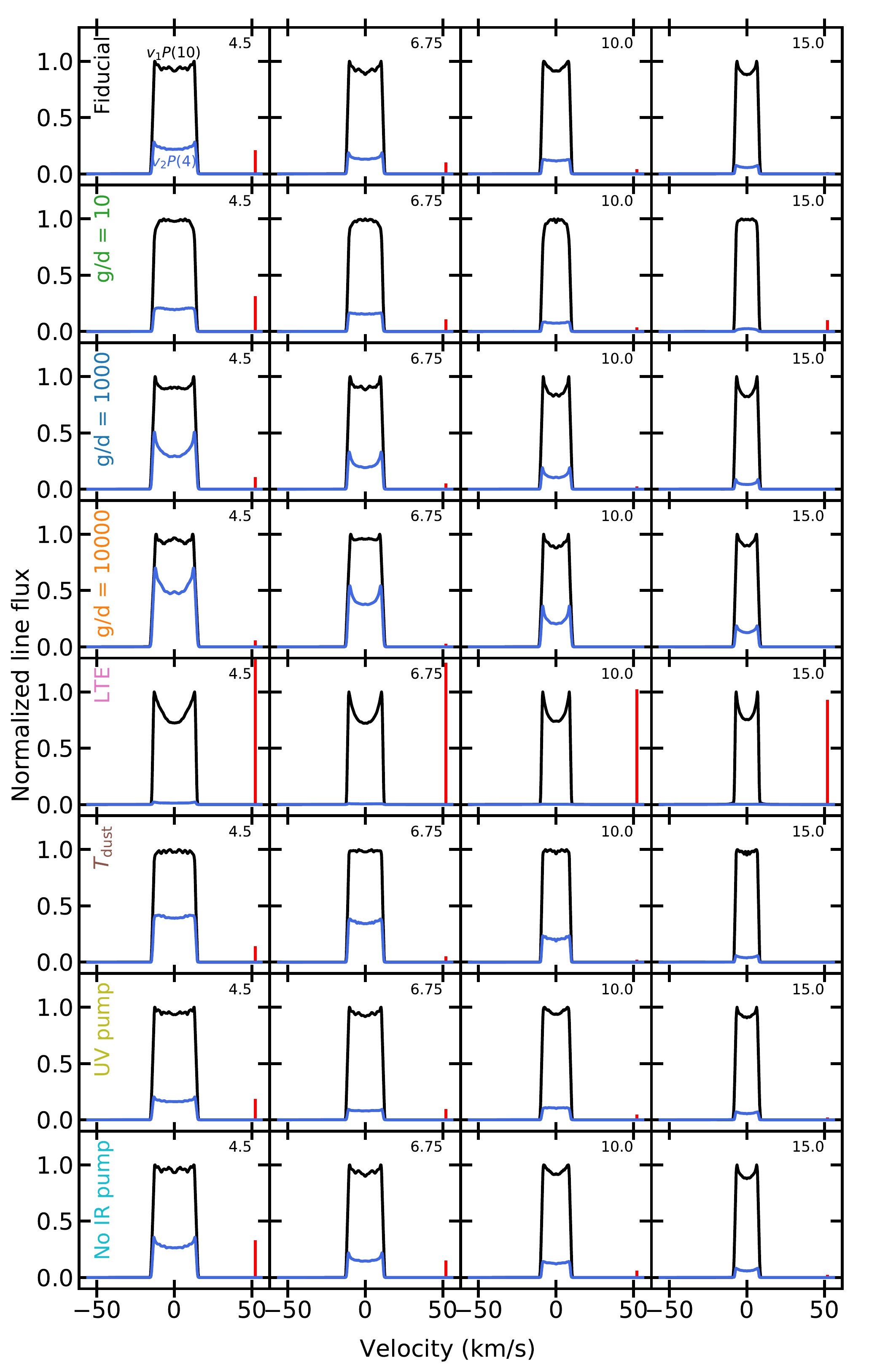}
\caption{\label{fig:lineprofiles_large} Same as Fig.~\ref{fig:lineprofiles}, but for models with inner radii $> 4.5$ AU. The right bar in the bottom right of each panel shows 0.3 times the continuum.}
\end{figure*}

Figure~\ref{fig:lineprofiles_large} shows the line profiles for the disks with larger holes. Almost all of the emission in these models is dominated by the inner wall. The lines are very strong with respect to the continuum, in some cases line-to-continuum ratios reach over 100. Part of this is the NIR excess, which is almost non existent, but even taking that into account, some of the line fluxes are a factor of 10 above the brightest lines observed.

\begin{figure*}
\includegraphics[width = 0.58\hsize]{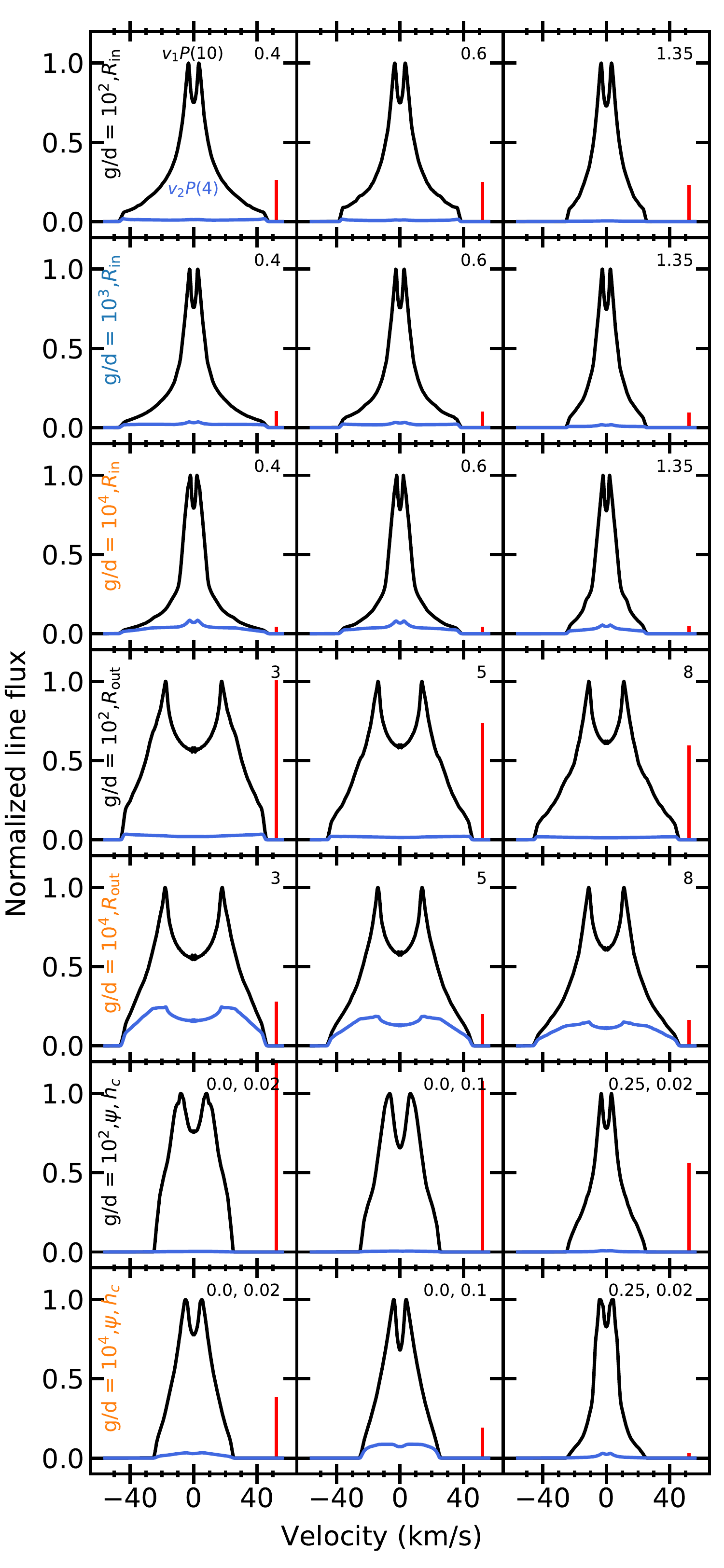}
\caption{\label{fig:lineprofiles_subtracted} Same as Fig.~\ref{fig:lineprofiles}, but for models were the inner rim contribution has been subtracted.}
\end{figure*}

Figure~\ref{fig:lineprofiles_subtracted} shows the line profiles for the disks with emission from the inner rim subtracted. Many of these models show very low $v2/v1$ emission. Variations in outer radius and vertical structure redistribute the CO emission over the disk surface, creating a large assortment of line profiles.

\section{Near-infrared excess}
\label{app:FNIR}
\begin{figure}
    \centering
    \includegraphics[width = \hsize]{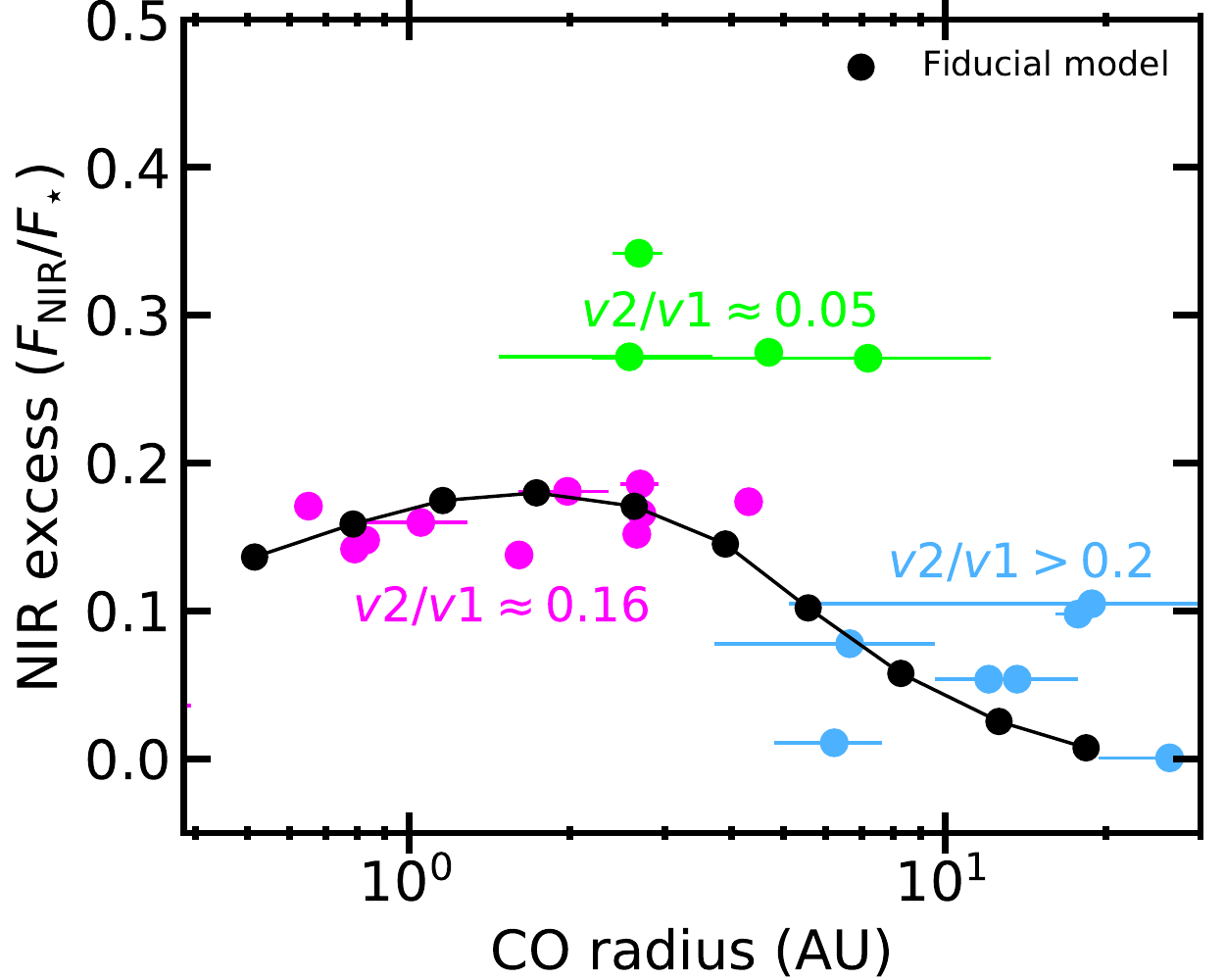}
    \caption{CO emitting radius versus the near-infrared excess flux as fraction of the total stellar flux. Black circles show the inner model radius against the near-infrared excess for the fiducial, g/d=100, model (see Sec.~\ref{sec:DALI} for details). The data are shown following the same color coding as in Fig.~\ref{fig:data}.}
    \label{fig:Fnir_RCO}
\end{figure}
The near-infrared continuum is another good probe of the inner disk. As all of the models discussed above have the same dust structure, they show identical near-infrared excesses, only depending on the inner edge radius of the disk in the model. Fig.~\ref{fig:Fnir_RCO} shows the near-infrared excess as function of $R_\mathrm{CO}$ for the gas-to-dust = 100 model and the data. The gas-to-dust 1000 and 10000 overlap almost exactly, the LTE and gas-to-dust 10 models deviate at small inner model radii. 

The models match the near-infrared excess range for most of the observations, except for the four observed disks with $F_\mathrm{NIR}/F_\star$ around 0.3. These disks have been shown to require very vertically extended dust structures to fit the NIR SED \citep{Maaskant2013}. The good match between the model and the observational near-infrared excesses indicate that the vertical extent, $h_c$, used in the model is reasonable for most of the observed sources. Up to $R_\mathrm{CO} = 5$ AU there is only very little variation in the NIR-excess in both the models and the group II objects, indication that up to 5 AU the NIR excess is a bad discriminator of gap size.

\subsection{CO as tracer of the inner disk radius}
\begin{figure}
    \centering
    \includegraphics[width = \hsize]{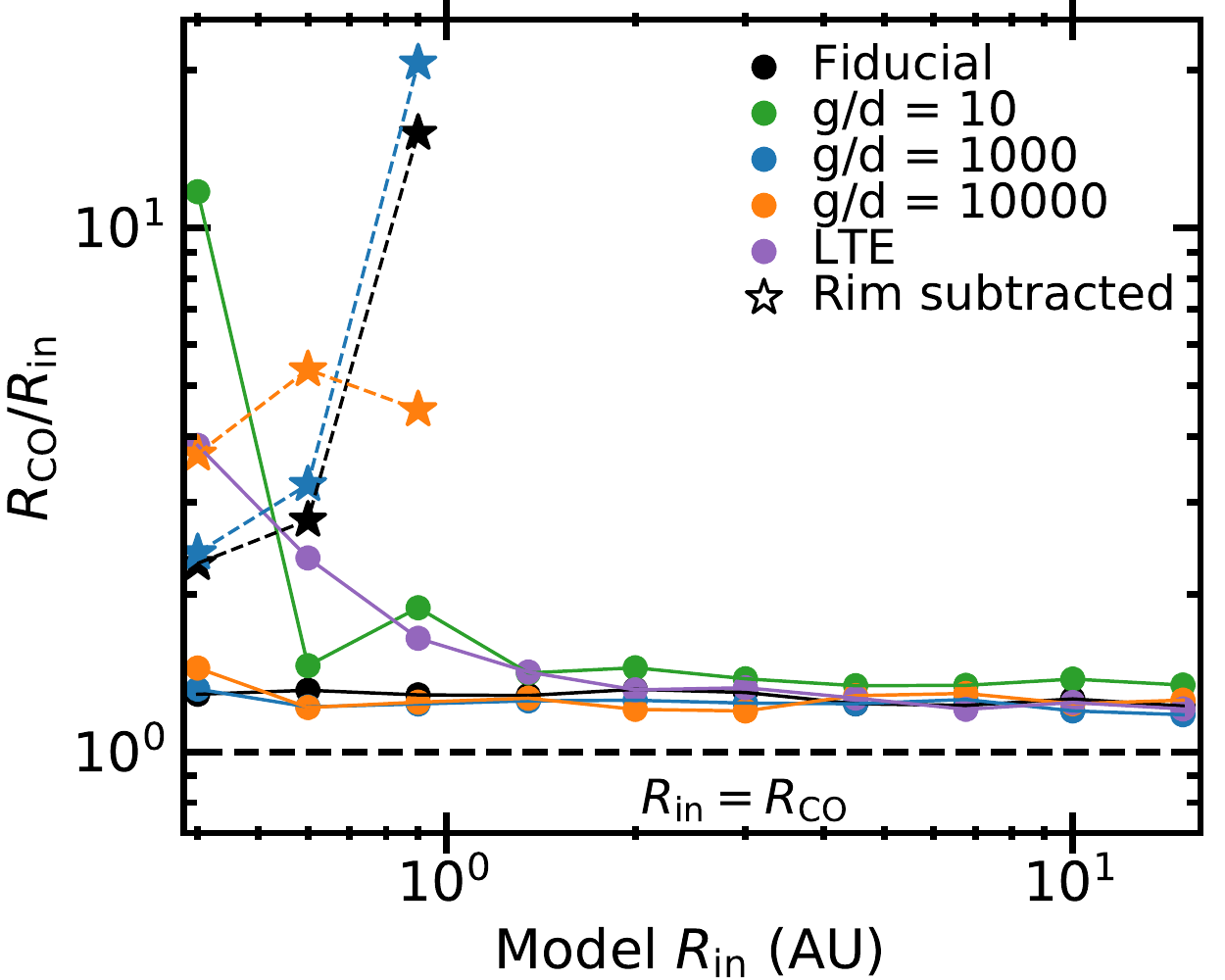}
    \caption{Model radius versus the deviation of $R_\mathrm{CO}$ from the inner model radius. Filled circles correspond to the normal DALI models, the stars correspond to the DALI models with the contribution of the inner rim removed. Only the models sets with varying inner rim location are shown. 
    }
    \label{fig:RCOvsRin}
\end{figure}

The half width half maximum of the $v1$ lines, or of their broad component where present (see Section \ref{sec:data}), is used to estimate a characteristic emitting radius for the inner molecular gas in the disk \citep[e.g.][]{Salyk2008, Brown2013, Banzatti2015}. Using the results from our thermo-chemical models we can infer how well the relation between CO emission radius and inner (molecular) disk radius holds. Figure~\ref{fig:RCOvsRin} shows that $R_\mathrm{CO}$ is always larger than the model inner radius. For the fiducial thermo-chemical models the deviations are around 50\%, if the inner rim of the disk is detected. If the inner rim is too weak to be detected, or in cases where it is artificially removed, the relation between $R_\mathrm{CO}$ and $R_\mathrm{in}$ breaks down as the disk surface far beyond the inner disk radius can dominate the emission.

The low vibrational ratios at small $R_\mathrm{CO}$ can only be matched by disk surface emission. It is thus for these sources that the largest discrepancy is expected between the measured $R_\mathrm{CO}$ and the innermost radius where dust or molecular gas is actually present. The model explorations in Sec.~\ref{sec:Sep_inner_rim} show that a single inner model radius can lead to a spread in measured CO radii. In these cases the line profiles give a clear sign of CO emission within $R_\mathrm{CO}$.

The high vibrational ratios at large $R_\mathrm{CO}$ are best matched by emission from a inner cavity wall. As such $R_\mathrm{CO}$ tightly traces the inner most radius at which CO is present. This is also clearly seen in the observed line profiles for these sources which have very steep sides.

\section{Lowering the flux of the outer disk}
\label{app:outerdisk}

\begin{figure}
    \centering
    \includegraphics[width = \hsize]{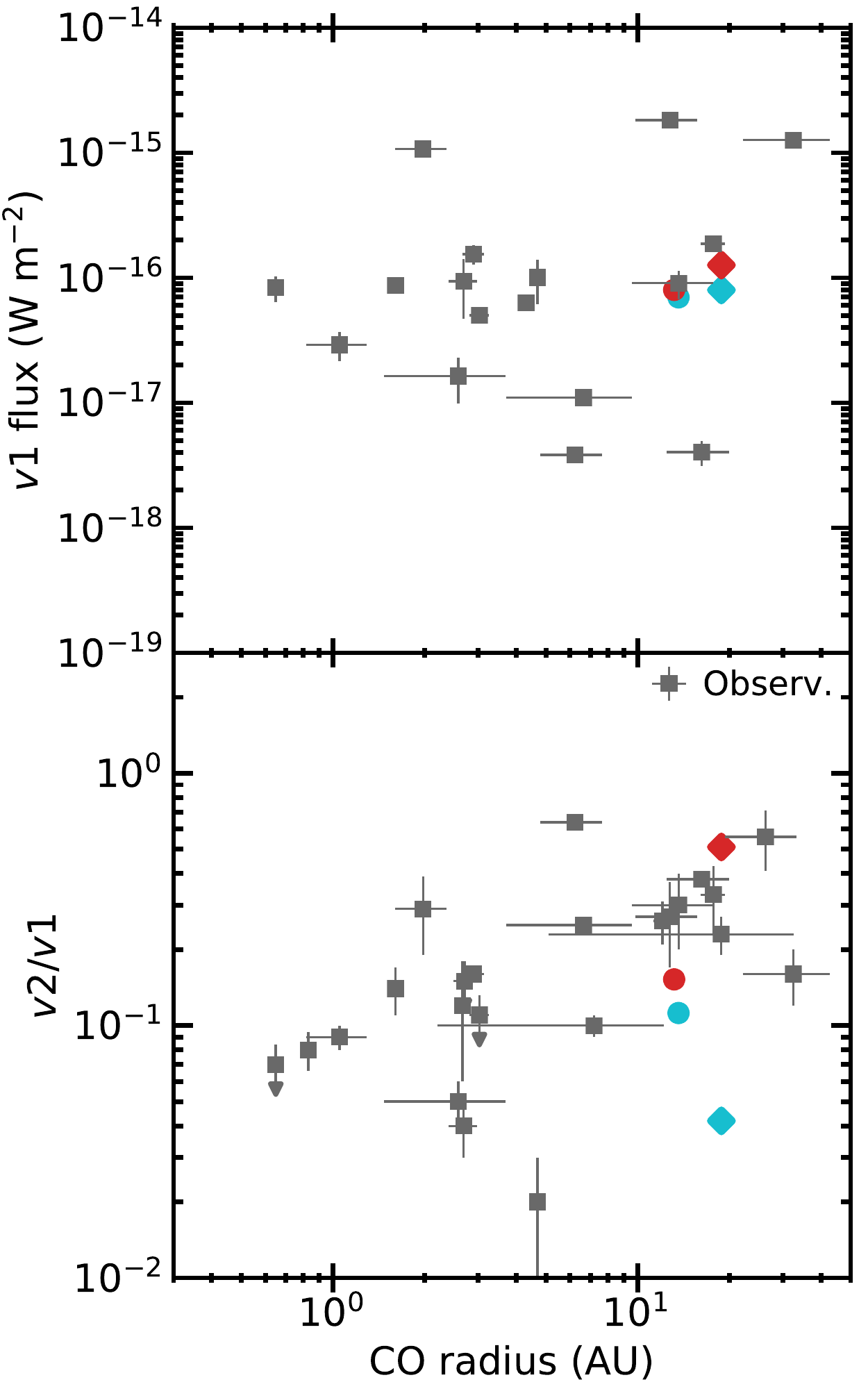}
    \caption{ $v1$ line flux (\textit{top}) and vibrational ratio of CO (\textit{bottom}) versus the inferred radius of emission for observational data and DALI model results. Cyan and red markers show models with a gas-to-dust ratio of 20000 and 100000 respectively. Circles show models with $R_\mathrm{in} =10$ AU, squares show models with with $R_\mathrm{in} =15$ AU }
    \label{fig:lowflux_highvib}
\end{figure}

The fiducial, high gas-to-dust ratio models that can match both $R_\mathrm{CO}$ and the high line ratios have $v1$ fluxes that are around a factor 50 higher than that of the average source. Comparisons between the RADEX model predictions in Fig.~\ref{fig:RADEX_flux_ratio_match} and the emitting area in the DALI models with 10 and 15 AU cavities showed that both the emitting area and the temperature of the gas had to be reduced. Reducing the models scale height from 0.1 to 0.03, in line with \citet{vanderMarel2016} lowered the fluxes by about a factor of 3 (e.g. HD 142527, HD 135344B,  HD 36112 and HD 31293). The emitting area was further reduced by putting $\chi_\mathrm{set} =1$. This lowered the flux from the disk surface. To lower the gas temperature at the inner wall of the disk we considered a different structure of the inner wall than the step-function we had. It turned out that a structure with an decreasing density inwards of the inner radius worked well. The gas density within the inner radius ($R_\mathrm{in}$) was parameterized by:
\begin{equation}
n_\mathrm{gas}(r,z) = n_\mathrm{gas}(R_{\mathrm{in}}, 0) \frac{1}{\sqrt{2 \pi} H(R_\mathrm{in})} \exp\left[-\frac12 \frac{\left(r-R_\mathrm{in}\right) ^ 2  + z^2}{H^2(R_\mathrm{in})}\right],
\end{equation}
where $n_\mathrm{gas}(R_{in}, 0)$ is the midplane density at the inner disk edge from the fiducial model. $H(r) = h_c r \left(\frac{r}{R_c}\right)^\psi$, with the parameters as defined in Table~\ref{tab:All_mod_param}. $R_\mathrm{in}$ was taken to be either 10 or 15 AU. Within the cavity radius, it was assumed that there were only small grains and the dust density was decreased to reach gas-to-dust ratios of 20000 and 100000. The CO lines from these models were calculated and are shown in Fig.~\ref{fig:lowflux_highvib}. 

The high gas-to-dust ratios are necessary to reach get the vibrational ratios to the observed values. At these gas-to-dust ratios the dust is nearly transparent to the CO lines and varying the gas-to-dust ratio has little effect on the line ratios. To further increase the line ratios, larger gas columns would be needed. 

These structures are probably not the only structures that can explain the CO rovibrational observables. Modelling of other observations of CO such as high resolution ALMA images or \textit{Herschel} observations of high $J$ CO lines would be able to differentiate between different structures. This modelling will have to be done on a source by source basis and is left to future work.

\section{Thermal dissociation of CO}
\label{app:thermdiss}
As mentioned before dissociating CO is difficult. The CO photo-dissociation rate is an order of magnitude lower than the dissociation rate for \ce{CO2} or \ce{H2O} for a 10000 K black body radiation field. Furthermore, most reactions destroying CO have a large barrier \citep[e.g. \ce{CO + H -> C + OH}, 77000K][]{Hollenbach1979,Mitchell1984} or create products that dissociate back into CO (e.g. \ce{CO  + OH -> CO2 + H}).

The strong UV fields near a Herbig star, especially when coupled with a low dust opacity, can photo-dissociate CO. This is most effective at lower densities. At higher densities ($10^{12}-10^{14}$ cm$^{-3}$), the chemical model in DALI shows that the CO production rate is faster than the UV radiation field expected at the sublimation radius. The high radiation field could however increase the gas temperature to far above the dust temperature. 

\begin{figure}
    \centering
    \includegraphics[width = \hsize]{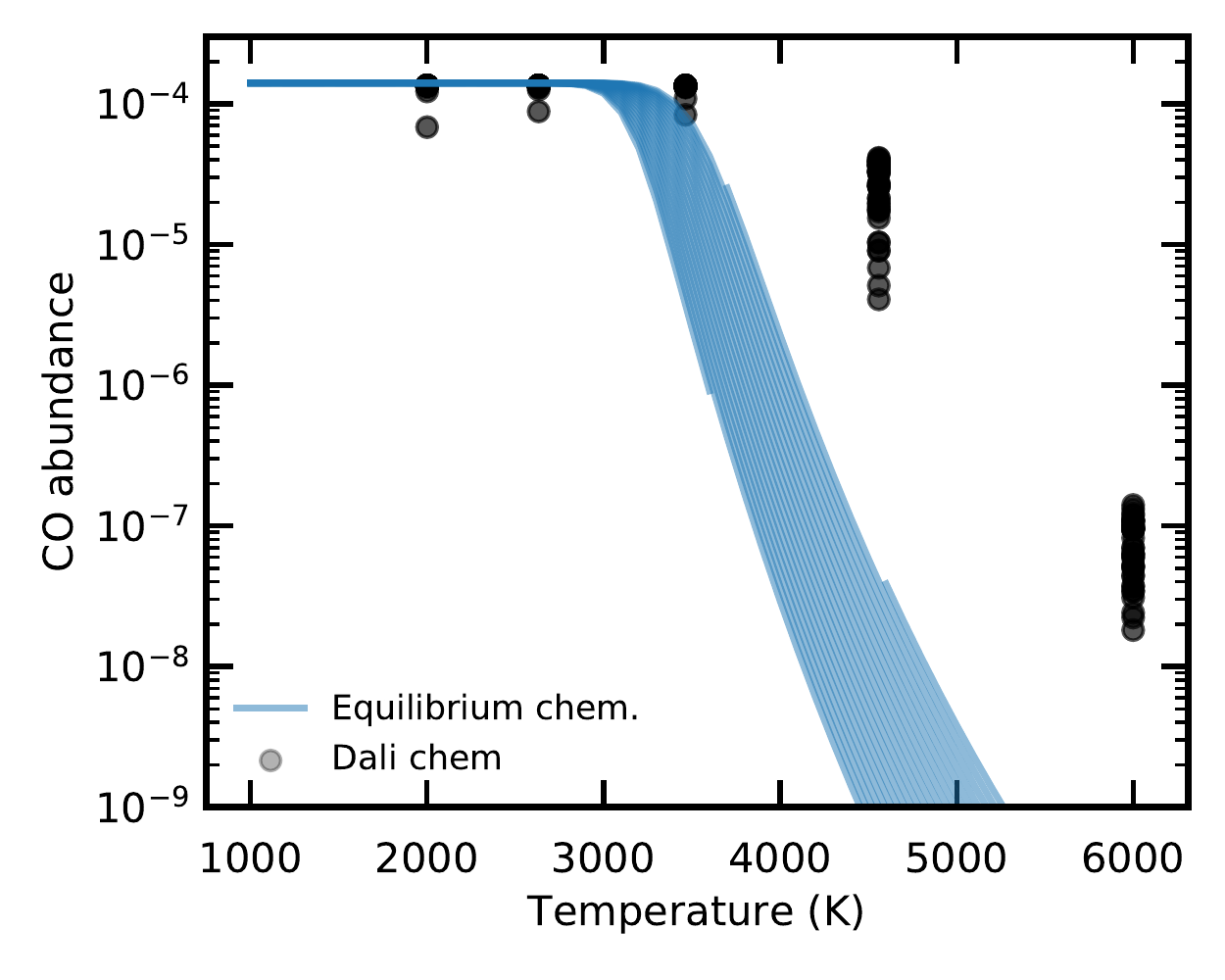}
    \caption{CO abundance as function of temperature under inner disk conditions. Black points show the results from the chemical model in DALI under conditions relevant for the inner disk ($n=10^{12}$--$10^{14}$ $cm^{-3}$, $10^4$--$10^{10} G_0$ radiation field). Blue line show the results from equilibrium chemistry for the same range of densities \citep[Chemical Equilibrium
with Applications,][]{Gordon1994,McBride1996}. }
    \label{fig:chem_comp}
\end{figure}

At temperatures above 3000 K Gibbs free energy minimisers find CO abundances lower than the canonical $10^{-4}$ (Fig.~\ref{fig:chem_comp}). Around this temperature, the kinetical network also shows a decrease in the CO abundance as the endothermic \ce{CO + H -> C + OH} and \ce{OH + H -> O + H2} reactions coupled with collisional dissociation of \ce{H2} push the gas towards a fully atomic state.

Figure~\ref{fig:chem_comp} clearly shows that the molecular to atomic transition happens slower and at higher temperatures than in the equilibrium models. The kinetic model does not have all the reaction pathways included that could be important for the production and destruction of CO and \ce{H2} at high temperatures. Between the uncertainty in the gas temperature and the uncertainty in the chemistry, it is thus not unlikely that the CO abundance in the inner disk rim is overestimated by DALI.

\end{document}